\documentclass[12pt,a4paper,english,fleqn]{article}
\usepackage[sort&compress,round,comma,authoryear]{natbib}
\usepackage[T1]{fontenc}
\usepackage{ae,aecompl}
\usepackage[colorlinks=true,urlcolor=blue,citecolor=blue,linkcolor=red,bookmarks=true]{hyperref}
\usepackage{graphicx}
\usepackage{float}	% Including figure files
\usepackage{amsfonts}
\usepackage{babel}
\newlength{\lp}
\makeatother
\begin{document}
\title{Conservative MOND from first principles.}
\author{D. F. Roscoe (The Open University; D.Roscoe@open.ac.uk)\\ \\ORCID: 0000-0003-3561-7425}
\date{}
\maketitle
\newpage
% Abstract of the paper
\begin{abstract}
The primary argument levelled against Milgrom's MOND is that it has no theoretical support, even though considerable effort has been expended in attempting to provide it. Against that criticism, MOND irrefutably enjoys an expanding portfolio of success and so is almost certainly tapping into something fundamental. But what?
\\\\
Over roughly the same period that MOND has been a topic of controversy, Baryshev, Sylos Labini and others have been claiming, with equal controversy in earlier years, that, on medium scales at least, material in the universe is distributed in a quasi-fractal $D\approx 2$ fashion. There is a link: if the idea of a quasi-fractal $D\approx 2$ universe on medium scales is taken seriously, then there is an associated characteristic mass surface density, $\Sigma_F$ say, and an associated characteristic acceleration scale $a_F = 4\pi G \,\Sigma_F$.
\\\\ 
The whole success of MOND is predicated upon the idea of a critical acceleration scale, $a_0$. It is an obvious step to make the association $a_0 \sim a_F$ and then to consider the MOND critical acceleration boundary simply as a marker for a characteristic mass surface density boundary separating \lq{galaxy}' from an environment characterized by $\Sigma_F$. This provides a route to the synthesis of conservative MOND from first principles.
\\\\
The radial acceleration relation (RAR) for conservative MOND when applied to the SPARC sample is the unity line. There is no mass discrepancy.
\end{abstract}
\newpage{}
% Select between one and six entries from the list of approved keywords.
% Don't make up new ones.
%\begin{keywords}
%MOND -- Fractal -- homogeneity
%\end{keywords}

%%%%%%%%%%%%%%%%%%%%%%%%%%%%%%%%%%%%%%%%%%%%%%%%%%

%%%%%%%%%%%%%%%%% BODY OF PAPER %%%%%%%%%%%%%%%%%%
\section{Introduction:}\label{Intro}
Modern ideas about mass-modeling within galaxies fall into one of two categories: 
\begin{itemize}
	\item the general concensus is that some form or other of Dark Matter is essential if the observed dynamics within (generally) spiral galaxies is ever to be understood;
	\item resisting the general concensus is the very much minority view that a modification of the classical Newtonian theory is required - a view that is encapsulated within the MOND algorithm, introduced by Milgrom in the 1980s.
\end{itemize}
Milgrom, along with several other authors over the years, puzzled over the dual mysteries of the \lq{flat rotation curve}' phenomenon of disk galaxies and the baryonic Tully-Fisher relationship which related the asymptotic (flat) rotation velocity in such galaxies to their visible mass. His crucial insight in the early 1980's was the recognition that if the flip to flatness of rotation curves occurred on an \emph{acceleration} scale, rather than some distance scale which many had tried, then the baryonic Tully-Fisher relationship would follow as a natural conequence. This idea, MOND, (for Modified Newtonian Dynamics) proved to be surprisingly productive, as is evidenced by the blitz of work which followed \citet{Milgrom1983a, Milgrom1983b,Milgrom1983c,Milgrom1983d,Milgrom1984,Milgrom1988,Milgrom1989a,Milgrom1989b,Milgrom1989c,Milgrom1991,Milgrom1994a,Milgrom1994b,Milgrom1995,Milgrom1997a,Milgrom1997b,Milgrom1998,Milgrom1999}. Sanders \citet{Sanders1984, Sanders1986, Sanders1988, Sanders1989, Sanders1990, Sanders1994a, Sanders1994b, Sanders1996, Sanders1997, Sanders1998a, Sanders1998b, Sanders1999, Sanders2000, Sanders2001, Sanders2014},  McGaugh  \citet{McGaugh1995a,McGaugh1995b,McGaugh1996,McGaugh1998a,McGaugh1998b,McGaugh1998c,McGaugh1999a,McGaugh1999b,McGaugh2000a, McGaugh2000b,McGaugh2001} (and others, variously)   added considerably to the volume of work demonstrating the absolute efficacy of the MOND algorithm in the context of disk galaxies. These references are inclusive up until about the turn of the century. In more recent times, MOND has been used with considerable success in the modelling of globular clusters, for example \citet{Kroupa1}, and  in the large hydrodynamics codes to model galaxy formation, \citet{Kroupa}. 
\\\\
The primary argument levelled against MOND (apart from the fact that its successes are so far confined to the domain of galaxies and clusters of galaxies) is that it has no theoretical support, although significant effort has been expended in trying to incorporate it, somehow or other, into General Relativity. 
\\\\
We develop conservative MOND (which is completely general in its applicability) and use the special case which refers to rotationally supported systems to analyse the SPARC sample of \citet{McGaugh2015}. Given complete dynamical information, it returns absolute distance scales, calculated independently of standard candles and the photometric method, and complete whole-disk mass profiles which map perfectly (in a statistical sense) onto SPARC photometric mass profile determinations. The RAR, computed using the SPARC sample, is the unity line. There is no mass discrepancy.
\\\\
All linear regressions in this paper are performed using a \emph{least areas} regression technique which treats predictor errors and response errors in a perfectly symmetric way. Thus, predictor and response can be interchanged leaving the predicted relation between the variables concerned invariant. See appendix \S\ref{LeastAreas}.
\section{Conservative MOND: Overview} 
Before expanding the details, we begin by giving a comprehensive overview of conservative MOND applied to the special case of rotationally supported systems.
\subsection{The Radial Acceleration Relationship} \label{RAR}
\begin{figure}[H]
	\centering
	\includegraphics[height=0.9\linewidth]{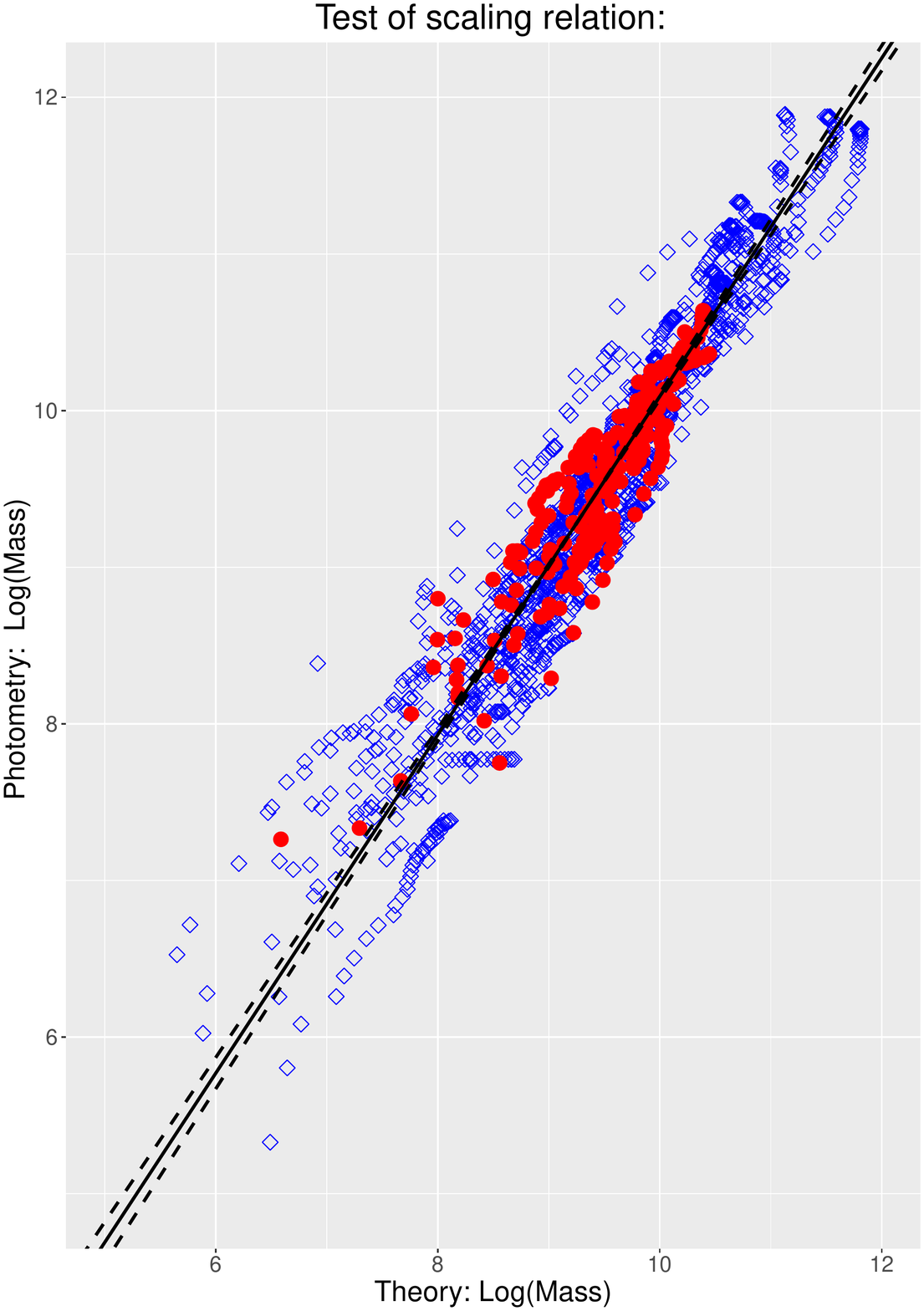}
	\caption{ Sample: 118 non-bulgy SPARC objects with an inclination $> 30^o$ giving a total of 1982 individual data points. Red filled circles = data from 25 designated LSB objects. Open diamonds = everything else.  Units are in $10^9$ solar masses. $MLR = 2.0$, globally.}
	\label{fig:RAR}
\end{figure}

The story of conservative MOND for rotationally supported systems is a story of scaling relationships (from one of which the RAR derives) normalized to the critical acceleration boundary.  Because of the central importance of the RAR concept in the contemporary discourse around MOND (\citet{McGaugh2016}), we give an introduction from the perspective of conservative MOND, fleshing out the details in the main text. The primary result here is that the rotation velocities are completely accounted for by the photometrically estimated mass for all objects in the sample including the LSBs. This is the essential message of figure \ref{fig:RAR}.
Applied to a rotationally supported system, conservative MOND provides the means (via a rotation curve fitting algorithm which is independent of centripetal acceleration observations) of estimating $(R_0, M_0, V_0)$, where $R_0$ is the theoretically determined critical acceleration radius, $M_0$ is the theoretically determined mass contained within $R_0$ and  $V_0$ is the rotation velocity at $R_0$, together with the scaling relations
\begin{eqnarray}
M(R)({\rm theory}) &=& M_0  \left( \frac{R}{R_0}\right)^2\left( \frac{V_0}{V_{rot}(R)}\right)^2,~~~ R  \leq R_0, \nonumber \\
\label{RAR1} \\
M(R)({\rm theory}) &=& M_0 \left( \frac{V_{rot}(R)}{V_0}\right)^2,~~~ R > R_0,~~V_{rot}(R) \leq V_{flat}
\nonumber
\end{eqnarray}
where $M(R)$(theory) is the predicted baryonic mass contained within radius $R$.
\\\\
At face value, these scaling relationships cannot possibly apply to LSB objects (of which there are at least 25 in the SPARC sample) since LSBs are defined as objects for which there is no critical boundary implying that $R_0=0$ and $M_0=0$ with $V_0$ being undefined. On this basis alone, we can expect (\ref{RAR1}) to be falsified. We find the very opposite to be true, as figure \ref{fig:RAR} so vividly shows.
\\\\
Briefly, this LSB conflict is resolved as follows: classical MOND locates the critical acceleration boundary in a disk using apparent centripetal accelerations, which depend on the photometric distance scale. By contrast, conservative MOND locates the critical acceleration boundary using a black-box rotation curve fitting algorithm, which is independent of the photometric distance scale, and then normalizes the distance scale for each object to ensure that $V_0^2/R_0 = a_0 \approx 1.2\times 10^{-10}\,mtrs/sec^2$ at the identified location of the critical boundary. In practice, this black-box process returns non-trivial $(R_0,M_0,V_0)$ for every LSB object in the sample; the scaling relations (\ref{RAR1}) then align those LSBs perfectly with the non-LSB objects in figure \ref{fig:RAR}. So, as a matter of objective fact, LSBs and non-LSBs behave identically under the scaling relations (\ref{RAR1}). The implications of all this are discussed in detail in the main text.
\\\\
Furnished with values for $(R_0,M_0,V_0)$ from the RC fitting algorithm of \S\ref{Details}, we then construct figure \ref{fig:RAR} by integrating the photometry for each disk in the SPARC sample (using a global $MLR = 2.0$, which is coarse but effective) to obtain values $M(R_i)(\rm{photometric})$ which are estimates of the baryonic mass contained within radius $R_i$, for each radial coordinate $R_i > 0$ on the measured disk. This gives a total of 1982 individual photometrically estimated mass values over the whole non-bulgy SPARC sample for disks with inclination $> 30^o$. Figure \ref{fig:RAR} then plots $M(R_i)(\rm{photometric})$ against  $M(R_i)(\rm{theory})$ defined by (\ref{RAR1}). It is manifestly the case that the LSB objects behave exactly as the non-LSB objects under the scaling relations (\ref{RAR1}).
\\\\
In figure \ref{fig:RAR} itself, the red-filled circles represent objects identified as LSBs in the source papers used by \citet{McGaugh2015}, whilst the open blue diamonds represent everything else. We find for all objects:
\[
\log M(R)({\rm photometric}) = (1.08\pm 0.03) \log M(R) ({\rm theory}) - (0.71 \pm 0.28) 
\]
whilst for the LSBs alone we find:
\[
\log M(R)({\rm photometric}) = (1.00\pm 0.05) \log M(R) ({\rm theory}) + (0.06 \pm 0.46). 
\]
It is clear, without further analysis, that the relationship
\[
M(R){\rm (theory)}  \approx M(R){\rm (photometric)} 
\]
is confirmed beyond doubt on the SPARC sample. The rotation velocities are completely accounted for by the photometrically determined baryonic mass. Putative LSB objects behave exactly the same under the analysis as do the non-LSB objects. There is no mass discrepancy.
\\\\
The RAR for conservative MOND follows directly from (\ref{RAR1}), and can be written as:
\begin{eqnarray}
\left[\frac{V_{rot}^2(R)}{R}\right]({\rm observed}) &=&  \left(\frac{V_0^2}{R_0} \right) \left(\frac{R}{R_0} \right) \left[ \frac{M_0}{M(R)}\right] \rm{(photometric)},~~~ R \leq R_0,
\nonumber \\
\nonumber \\
\left[\frac{V_{rot}^2(R)}{R}\right]({\rm observed}) &=& \frac{V_0^2}{R} \left[ \frac{M(R)}{M_0} \right]({\rm photometric}),~~~R > R_0, ~~~ V_{rot}(R) \leq V_{flat}. \nonumber
\end{eqnarray}
where $(R_0,M_0,V_0)$ are the values returned by the RC fitting algorithm for each object and $M(R)$ is determined directly from SPARC photometry. In an obvious notation, we find:
\[
g{\rm (observed)} = \left(1.02 \pm 0.04 \right) g{\rm (baryonic)} - \left(0.06 \pm 0.14 \right)
\]
so that, as figure \ref{fig:RAR} implies, the RAR for conservative MOND follows the unity line. 
\subsection{Summary of the theory} \label{OverView}
The foundation for conservative MOND is a fully conservative cosmology, rooted in the Leibniz-Mach worldview, synthesized from an intensive analysis of how clocks and rods are to be defined in that worldview. 
\\\\
The synthesis appeared in the mainstream literature in an early form as \citet{Roscoe2002}. This early form is mathematically complete, but only partially interpreted around the meaning of clocks in the equilibrium state. The fully interpreted form is given as online supplementary material here, and exists in the archive as \citet{Roscoe2020}. A brief outline of the technical details is given in \S\ref{LMC}.
\\\\
The fundamental object, $\mathcal{M}(\Phi)$, from which the cosmology flows is defined as follows:
\begin{itemize}
	\item The level surface $\Phi=k$: in the most simple case this represents a three-dimensional spherical surface but, in general, it represents any topological isomorphism of a three-dimensional spherical surface. For example, a thick pancake;
	\item The normalized mass function $\mathcal{M}(\Phi)$: this represents the total mass contained within the level surface $\Phi=k$ with any mass singularity at the origin removed, so that $\mathcal{M}(0)=0$.
\end{itemize}
The irreducible basic solution of the Leibniz-Mach cosmology is an equilibrium world:  material is necessarily non-trivially present in the form of a $D=2$ fractal distribution, and all material motions are unaccelerated. Consquently, about any point chosen as the centre, the geometry can be considered as spherical and mass is distributed according to
\begin{equation}
\mathcal{M}(R) = 4 \pi \Sigma_F R^2,  \label{Frac1}
\end{equation}
about the chosen centre, and where $\Sigma_F$ is the characteristic mass surface density of the distribution.
This is in  direct conformity with the modern perspective that, on medium scales at least, material in the universe is distributed in a quasi-fractal $D\approx 2$ fashion. A full discussion of this point is given in \S\ref{GeneralConsiderations}.
\\\\
A galactic system is then represented as a bounded perturbation of (\ref{Frac1}). For the simple case of a purely spherical distribution, it has the form:
\begin{eqnarray}
\mathcal{M}(R) &\equiv& \mathcal{M}_g(R),~~~~ R \leq R_0; \nonumber \\
\nonumber \\
\mathcal{M}(R) &\equiv&  \mathcal{M}_g(R_0)  + 4 \pi \Sigma_F (R^2-R_0^2),~~~~ R > R_0; \label{Frac2} \\
\nonumber \\
\mathcal{M}'_g(R_0) &=& 8 \pi \Sigma_F R_0, ~~~\rm{(Gradient~continuity~condition)};\nonumber
\end{eqnarray}
where:
\begin{itemize}
	\item $\mathcal{M}_g(R)$ represents the galactic mass contained within radius $R\leq R_0$;
	\item $R_0$ is the unspecified radial boundary of the perturbation;
	\item The gradient continuity condition (\ref{Frac2}c) at the boundary between the two regimes is required (in the case of arbitrary orbits) to ensure that the system potential, $\mathcal{V}(R)$, which defines the system clock, is uniquely defined at the boundary $R=R_0$. This condition leads directly to the definition of this boundary as a critical acceleration boundary where the critical acceleration $\sim 8\pi G \Sigma_F$.
\end{itemize}
{\bf Conservative MOND} is then simply the Leibniz-Mach cosmology which flows from the mass model (\ref{Frac2}), or the generalization of that mass model to any topological isomorphism of a spherical distribution; for example, the thick pancake.
\subsection{Classical MOND critical acceleration condition}
For convenience, write the mass model on $R\leq R_0$ as
\begin{equation}
\mathcal{M}_g(R) \equiv M_0 \Psi\left(\frac{R}{R_0} \right),~~~~R \leq R_0,~~~ \Psi(1) = 1 \label{Frac5}
\end{equation}
where $\Psi(X)$ is a differentiable function of $X\equiv R/R_0$.
Defining  $a_F\equiv 4 \pi G \Sigma_F$ as the characteristic acceleration associated with $\Sigma_F$,
then condition (\ref{Frac2}c) gives directly
\begin{equation}
\frac{M_0 G}{R_0^2} \, \Psi'(1) =  8 \pi G\, \Sigma_F \equiv 2 a_F. \label{Frac7}
\end{equation}
The foregoing relation applies to arbitrary orbits in spherical distributions. If we assume Newtonian gravitation at the critical boundary (not strictly true here, but the ansatz serves to make the point) then, for the case of purely circular motions, we have \emph{the MOND acceleration condition}:
\begin{equation}
\frac{M_0G}{R_0^2} = \frac{V_0^2}{R_0} = \frac{2 a_F}{\Psi'(1)}  \label{Frac8}
\end{equation}
to give the result of classical MOND relating the centripetal acceleration on the critical boundary to a global characteristic acceleration. The difference between classical MOND and conservative MOND is simply that whereas for classical MOND, $R=R_0$ is the boundary between two different laws of gravitation, for conservative MOND it is simply the boundary between the galaxy and its external environment.
\\\\
There is an obvious consequence of this simple analysis: by (\ref{Frac7}) (which comes from (\ref{Frac2}c)), the galactic object does not merely sit within its environment, but is actively tied into it in a way which ensures that the system clock (defined by the system potential) is uniquely defined at the critical boundary.
\subsection{Summary of SPARC analysis} \label{Overview}
The most simple possible realization of conservative MOND is that which admits circular motions only (rotationally supported systems), and this arises as a degenerate state of the radial Euler-Lagrange equation, (\ref{(11)}), leading to the scaling relation
\begin{equation}
\frac{V_{rot}(R)}{V_{flat}} =  
\left( \frac{\Sigma_F}{\Sigma_R} \right)^{1/2},~~~~R > 0 \label{Frac4}
\end{equation}
where $\Sigma_R$ is the mass surface density at radius $R$, calculated from the mass model (\ref{Frac2}) on the two partitions, $R\leq R_0$ and $R>R_0$, and it is this which we use to analyse the SPARC sample. A summary of the results arising from degenerate state conservative MOND is given in the following.
\\\\
Briefly, using the mass model (\ref{Frac2}), the general scaling relation (\ref{Frac4}) can be written as two scaling relations:
\begin{eqnarray}
S_1(R,V_{rot}(R),\Sigma_R,\Sigma_0,\Sigma_F,V_{flat})&=&0,~~~~~R\leq R_0; \nonumber \\
S_2(R,V_{rot}(R),\Sigma_R,\Sigma_0,\Sigma_F,V_{flat})&=&0,~~~~~R > R_0 \nonumber
\end{eqnarray}
where the additional parameter $\Sigma_0$ is the mass surface density at $R_0$. 
\\\\
The results flowing from this two-state scaling relationship can be listed as:
\begin{itemize}
	\item $S_1(R_0...)=0$ is a quantitative statement of Freeman's Law; 
	\item the BTFR emerges directly when the condition $S_1(R_0...)=0$ is further constrained by the MOND acceleration condition, $V_0^2/R_0=a_0$;
	\item in \S\ref{SPARCmass1} it is shown how the radial distribution of mass on the interior, $M(R \leq R_0)$(theory), tracks SPARC photometry with perfect statistical fidelity, using a global $MLR=2.0$;
	\item in \S\ref{SPARCmass2} it is shown how the radial distribution of mass on the exterior, $M(R > R_0)$(theory) tracks SPARC photometry up to $M_{flat}$ with perfect statistical fidelity, using a global $MLR=2.0$;
	\item all objects, including the 25 designated LSBs, in the SPARC sample are treated identically and with equal success;
	\item in \S\ref{RCbehaviour-1} it is shown how the scaling relationship $S_2(R>R_0)=0$ provides a perfect understanding of why RCs, having reached $(R_0,V_0)$ continue by:
	\begin{itemize}
	 \item either rising smoothly on an asymptotic approach to a rotation velocity $V_{flat} > V_0$;
	 \item or changing abruptly to flatness with a rotation velocity $V_{flat} = V_0$; 
	 \item or falling smoothly on an asymptotic approach to a rotation velocity $V_{flat}< V_0$.
	 \end{itemize}
	 In other words, degenerate case conservative MOND covers all observed rotation curve shapes in disk galaxies. 
\end{itemize}
\section{General considerations around conservative MOND} \label{GeneralConsiderations}
The irreducible basic solution of conservative MOND, corresponding to the trivial case of $R_0 \equiv 0$ in (\ref{Frac2}), is a state of equilibrium in which all motions are unaccelerated and where matter is distributed fractally, $D=2$.
\\\\
This theoretical result is in accordance with the now accepted reality is that, \emph{on medium scales at least}, matter in the universe is, in a statistical sense, distributed in a  $D\approx2$ quasi-fractal manner.  For this reason, \S\ref{Observations} provides a brief overview of the historical debate surrounding questions of large scale structure.
\\\\
However, there is a further implication which must be considered: the equilibrium solution of conservative MOND  implies idealized fractal $D=2$ behaviour over all scales. In practice, we interpret this to mean quasi-fractal behaviour extending to some undetermined minimum level of physical scale.
\\\\
Whatever this scale is,  the model (\ref{Frac2}) explicitly requires the existence of a non-trivial $D\approx 2$ quasi-fractal intergalactic medium (IGM) having exactly the same properties as the large scale distribution.
This raises an obvious question: if such an IGM exists, why has it not been detected? The answer to this question depends on the considerations of \S\ref{Observations}  and is addressed in \S\ref{Distances1} \& \S\ref{Distances2}.
\subsection{The external environment: the observations \& the debate}\label{Observations}
A basic assumption of the \textit{Standard Model}
of modern cosmology is that, on some scale, the universe is homogeneous;
however, in early responses to suspicions that the accruing data was
more consistent with Charlier's conceptions \citet{Charlier1908, Charlier1922, Charlier1924}
of an hierarchical universe than with the requirements of the \textit{Standard
	Model},  \citet{De Vaucouleurs1970} showed that, within wide limits,
the available data satisfied a mass distribution law $M\approx r^{1.3}$,
whilst \citet{Peebles1980} found $M\approx r^{1.23}$. The situation,
from the point of view of the \textit{Standard Model}, continued to
deteriorate with the growth of the data-base to the point that, \citet{Baryshev1995}  were able to say
\begin{quote}
	\emph{...the scale of the largest inhomogeneities (discovered to date)
		is comparable with the extent of the surveys, so that the largest
		known structures are limited by the boundaries of the survey in which
		they are detected.}  
\end{quote}
For example, several redshift surveys of the late 20th century, such
as those performed by \citet{Huchra1983}, \citet{Giovanelli1986}, \citet{DeLapparent1988}, \citet{Broadhurst}, \citet{DaCosta} and \citet{Vettolani} 
etc discovered massive structures such as sheets, filaments, superclusters
and voids, and showed that large structures are common features of
the observable universe; the most significant conclusion drawn from
all of these surveys was that the scale of the largest inhomogeneities
observed in the samples was comparable with the spatial extent of
those surveys themselves.\\
\\
In the closing years of the century, several quantitative analyses
of both pencil-beam and wide-angle surveys of galaxy distributions
were performed: three examples are given by \citet{Joyce}  who analysed the CfA2-South catalogue to find fractal
behaviour with $D\,$=$\,1.9\pm0.1$; \citet{SylosLabini}
analysed the APM-Stromlo survey to find fractal behaviour with $D\,$=$\,2.1\pm0.1$,
whilst \citet{SylosLabini1} analysed the
Perseus-Pisces survey to find fractal behaviour with $D\,$=$\,2.0\pm0.1$.
There are many other papers of this nature, and of the same period,
in the literature all supporting the view that, out to $30-40h^{-1}Mpc$
at least, galaxy distributions appeared to be consistent with the simple stochastic fractal model with the critical fractal dimension of $D\approx  D_{crit} = 2$.\\
\\
This latter view became widely accepted (for example, see  \citet{Wu}), and the open question became whether or not
there was transition to homogeneity on some sufficiently large scale.
For example, \citet{Scaramella} analyse the ESO Slice
Project redshift survey, whilst \citet{Martinez} analyse
the Perseus-Pisces, the APM-Stromlo and the 1.2-Jy IRAS redshift surveys,
with both groups claiming to find evidence for a cross-over to homogeneity
at large scales.\\
\\
At around about this time, the argument reduced to a question of
statistics (\citet{Labini}, \citet{Gabrielli}, \citet{Pietronero}):
basically, the proponents of the fractal view began to argue that
the statistical tools (that is, two-point correlation function methods) widely used
to analyse galaxy distributions by the proponents of the opposite
view are deeply rooted in classical ideas of statistics and implicitly
assume that the distributions from which samples are drawn are homogeneous
in the first place.  \citet{Hogg}, having accepted
these arguments, applied the techniques argued for by the pro-fractal
community (which use the \emph{conditional density} as an appropriate
statistic) to a sample drawn from Release Four of the Sloan Digital
Sky Survey. They claimed that the application of these methods does
show a turnover to homogeneity at the largest scales thereby closing,
as they see it, the argument. In response, \citet{SylosLabini2} 
criticized their paper on the basis that the strength of the
conclusions drawn is unwarrented given the deficencies of the sample
- in effect, that it is not big enough. 
\\\\
More recently, \citet{Tekhanovich} have addressed the deficencies of the Hogg et al analysis by analysing the 2MRS catalogue, which provides redshifts of over 43,000 objects out to about 300Mpc, using conditional density methods; their analysis shows that the distribution of objects in the 2MRS catalogue is consistent with the simple stochastic fractal model with the critical fractal dimension of $D\approx  D_{crit} = 2$.
\\\\
To summarize, the proponents of non-trivially fractal large-scale
structure have won the argument out to medium distances and the controversy
now revolves around the largest scales encompassed by the SDSS.
\subsection{General properties of a $D\approx 2$ quasi-fractal IGM}\label{Distances1}
Because of a perceived and unexplained over-dimming of type-SN1a supernovae, it has been hypothesized for some considerable time that there exists a non-trivial intergalactic medium (IGM) consisting of \lq{grey dust}' (dust which causes extinction, but very little reddening) expelled from the galaxies, to account for this. For example, see \citet{AA1999} and  \citet{PSC2006}. The principle of a \lq{grey dust}' IGM in some form or other is thus established, albeit as a means of explaining anomolous dimming of SN1a objects at high redshifts. Because of the high redshifts involved here, the inherent assumption is that the grey dust involved exists at extremely low column densities.
\\\\
However,  we can note that whilst the various proposed models assume some degree of homogeneity in the distribution of grey dust, the considerations of \S\ref{Observations} suggest otherwise: specifically, since galaxies on the medium distance scale are observed to be distributed in a $D\approx 2$ quasi-fractal manner, then so must any grey dust which originates in them be likewise distributed.
By virtue of its $D\approx 2$ quasi-fractal distribution, such an IGM would, to a significant extent, be transparent to radiation, which mirrors the primary reason why \citet{Charlier1908} suggested  the \lq{hierarchical universe}' as an early answer to the question \emph{Why is the sky dark at night?}
\\\\
To complicate the issue even further, we can reasonably suppose that this grey dust (should it exist) would be in thermodynamic equilibrium with the cosmic background, and therefore difficult, if not impossible, to distinguish from it.
The net result of these considerations is that an IGM consisting of a $D\approx 2$ quasi-fractal distribution of grey dust would  be extremely difficult to detect directly.
\subsection{Observational consequences of a  $D\approx 2$ fractal grey dust IGM} \label{Distances2}
Whilst a $D\approx 2$ quasi-fractal grey dust IGM would, to a significant extent, be transparent to radiation,
it would by no means be totally transparent and, broadly speaking, light from a source at distance $R$ would dim (without reddening) via a process of extinction by the grey dust in a way which would be proportional to $R^2$, again because this material is distributed quasi-fractally, $D\approx 2$. 
\\\\
In the absence of reddening, this dimming mechanism would be indistinguishable in its effects from the ordinary inverse-square distance dimming process so that the total of observed dimming would be interpreted entirely as a distance effect. There are two consequences:
\begin{itemize}
	\item The principles underlying the process by which standard candles are used to estimate the absolute luminosities of distant objects are unchanged so that such estimates would not be affected by grey dust extinction, should the phenomenon actually exist;
	\item The photometric distance scale would be systematically exaggerated with the effect that objects of a given absolute luminosity would generally be estimated as being further away and larger than they actually are. We can see that, in the context of classical gravitational theory, such an exaggeration of the distance scale would automatically give rise to a \lq{missing mass}' problem.
\end{itemize}
It is to be emphasized that it would be extremely difficult, if not impossible, to detect such a grey dust IGM by any direct means. Any dimming arising from its presence would be interpreted entirely as a distance effect.
\section{Degenerate case conservative MOND}\label{Outline}
According to the general theory (implicit to \citet{Roscoe2002}, but explicitly stated in \citet{Roscoe2020}) and shown at (\ref{(11)}), the dynamics associated with an arbitrary spherical $\mathcal{M}(R)$ admit a degenerate state in which only circular motions can occur, for which the radial Euler-Lagrange equation integrates (with a change of notation) to give:
\begin{equation}
V^2_{rot}(R) \mathcal{M}(R)  - V^2_{flat} 4 \pi R^2 \,\Sigma_F = 0. \label{eqn4g}
\end{equation}
For the case of general orbits, the condition (\ref{Frac2}c) requires that $\mathcal{M}'(R_0)$ is uniquely defined in order that the system clock was also uniquely defined at $R=R_0$. However, the degenerate case being considered here requires only that $\mathcal{M}(R_0)$ of (\ref{eqn4g}) is uniquely defined in order to give a uniquely defined system clock. The required continuity is already guaranteed by (\ref{Frac2}a) and (\ref{Frac2}b).
\\\\
Equation (\ref{eqn4g}) rearranges as the scaling relationship
\begin{equation}
\frac{V_{rot}(R)}{V_{flat}} =  \left( \frac{4 \pi R^2\, \Sigma_F }{\mathcal{M}(R)} \right)^{1/2}  \label{eqn4}
\end{equation}
from which, using (\ref{Frac2}), we see that $V_{rot} \rightarrow V_{flat}$ as $R \rightarrow \infty$ so that $V_{flat}$ is an asymptotic flat rotation velocity.
\\\\
The scaling relationship  (\ref{eqn4})  together with the mass model  (\ref{Frac2}) (without the gradient continuity condition)  gives degenerate case conservative MOND:
\begin{eqnarray}
\frac{V_{rot}(R)}{V_{flat}} &=& \left(\frac{ \Sigma_F }{   \Sigma_R } \right)^{1/2}, ~~~~ R > 0; \nonumber \\  
\nonumber \\
\Sigma_R &\equiv& \frac{\mathcal{M}_g(R)}{4 \pi R^2},~~~~ R \leq R_0; \label{eqn2} \\
\nonumber \\
\Sigma_R &\equiv& \frac{ \mathcal{M}_g(R_0)  + 4 \pi (R^2-R_0^2)\,\Sigma_F}{4 \pi R^2},~~~~ R > R_0; \nonumber
\end{eqnarray}
To summarize: writing $M_0\equiv \mathcal{M}_g(R_0)$, the primary disk parameters in degenerate case conservative MOND are $(R_0, M_0, V_{flat})$ where:
\begin{itemize}
	\item $\Sigma_R $ is the mass surface density at radius $R > 0$; 
	\item $\mathcal{M}_g(R\leq R_0)$ represents the radial distribution of mass within the critical radius;
	\item $M_0$ represents the total mass contained with $R_0$, the critical radius;
	\item $V_{flat}$ is the flat rotation velocity;
	\item the uniqueness of the system clock at $R=R_0$ is guaranteed by the continuity of $\mathcal{M}(R)$ at $R=R_0$.
\end{itemize} 
\section{Scaling relations} \label{BTFR}
In the following, we derive a quantitative refinement of Freeman's Law, the baryonic Tully-Fisher relation (BTFR) and the whole-disk scaling relationship for mass given at (\ref{RAR1}) and used in \S\ref{RAR} to demonstrate the RAR for conservative MOND. 
%\textcolor{blue}{
\subsection{The refined Freeman's Law } \label{FL-1}
From (\ref{eqn2}), at the critical boundary, $R=R_0$, we have immediately 
\begin{equation}
\left(\frac{V_0}{V_{flat}}\right)^2 = \frac{\Sigma_F}{\Sigma_0}  \label{eqn4e}
\end{equation}
from which we get an obvious refinement of Freeman's Law 
\begin{equation}
\Sigma_0 =\left( \frac{V_{flat}}{V_0}\right)^2  \Sigma_F  \label{eqn4f}
\end{equation}
relating the mass surface density of the galaxy at the critical radius to the characteristic mass surface density of the $D\approx 2$ fractal grey dust IGM.
\\\\
This general form makes it clear that the classical statement of Freeman's Law is restricted to those objects for which $V_0=V_{flat}$; that is, to those objects for which the critical acceleration, $a_0$, coincides with an abrupt transition to flatness. Such objects are those with architypal flat rotation curves - within the SPARC sample, two clear examples of such objects are ESO563-G02 and  NGC2998. 
\subsection{The BTFR} \label{BTFR-1}
From (\ref{eqn4e}) and using $a_F \equiv 4 \pi G \Sigma_F$,  we have 
\begin{equation}
V_0^2 = V_{flat}^2 \,\frac{\Sigma_F}{\Sigma_0} =  V_{flat}^2 \left( \frac{ a_F  R_0^2}{G M_0}\right)  \label{Frac6}
\end{equation}
where $M_0 \equiv \mathcal{M}_g(R_0)$ is the mass contained within the critical boundary.
\\\\
We have no direct information about the value of $a_F$, so we use the ansatz that the magnitude of the centripetal acceleration at $R=R_0$ is given by the MOND value of $a_0 \approx 1.2 \times 10^{-10}\,mtrs/sec^2$ together with the identification $a_F \equiv a_0$.
This ansatz allows us to write (\ref{Frac6}) as the two equations: 
\begin{equation}
\frac{V_0^2}{R_0} = a_0,~~~~  V_{flat}^2 \left( \frac{ R_0}{G M_0}\right) = 1. \label{eqn3NN}
\end{equation}
For the avoidance of confusion, and to emphasize the non-Newtonian nature of conservative MOND, note how the second of the two equations above differs from the Newtonian requirement that
\[
V_0^2 = \frac{G M_0}{R_0}.
\]
It is this difference which leads to the BTFR since eliminating $R_0$ between the two equations of (\ref{eqn3NN}) gives directly:
\begin{equation}
V_{flat}^4 = a_0\, G\, \left[ \left(\frac{V_{flat}}{V_0} \right)^2 M_0 \right]  \label{eqn4d}
\end{equation}
Defining $M_{flat}$(theory) according to
\begin{equation}
M_{flat}{\rm(theory)} \equiv \left(\frac{V_{flat}}{V_0} \right)^2 M_0, \label{eqn5d}
\end{equation}
then (\ref{eqn4d}) becomes
\begin{equation}
V_{flat}^4 = a_0\, G\, M_{flat}{\rm(theory)}  \label{eqn5c}
\end{equation}
which has the exact structure of Milgrom's form of the empirical BTFR. So everything hinges on the extent to which $M_{flat}{\rm(theory)}$ tracks $M_{flat}$(photometry), where we remember that this latter object is conventionally defined as the photometrically estimated mass contained up to $V_{flat}$. In practice, as is shown in \S\ref{Mflat},  we find $M_{flat}{\rm(theory)} = M_{flat}{\rm(photometry)}$ at the level of statistical certainty over the SPARC sample.
\\\\
However, notwithstanding the quality of this statistical result, it is clear that in those few cases for which $V_0>V_{flat}$, then  (\ref{eqn5d}) explicitly states that $M_{flat}{\rm(theory)} < M_0$. Consequently, since it is certainly true that $M_0<M_{flat}$(photometry), then  $M_{flat}{\rm(theory)} = M_{flat}{\rm(photometry)}$ cannot be true on the conventional definition of $M_{flat}{\rm(photometry)}$ in these few cases. The questions raised by this result are resolved in detail in the following,  \S\ref{SpecialCase-1}.
\subsection{The special case: $V_0 > V_{flat}$} \label{SpecialCase-1}
This is a minority case for the objects of the SPARC sample and crucially by (\ref{eqn4f}) (the refined Freeman's Law) is associated with the mass surface density condition $\Sigma_0 < \Sigma_F$. The considerations of \S\ref{RCbehaviour-1} show  that the RCs of such objects are characterized by a smooth \emph{descent} from $V_{max} \equiv V_0 > V_{flat}$ to approach $V_{flat}$ asymptotically.
\\\\
This case creates an interesting question in that, according to (\ref{eqn5d}), it implies $M_{flat}{\rm(theory)} < M_0$
so that $M_{flat}$ cannot be the contained luminous mass at the asymptotic $V_{flat}$ rotation velocity, contrary to the photometric definition. However, in such cases, the velocity value $V_{flat}$ is reached \emph{twice} on the rotation curve - once on the rising part before $V_{max}\equiv V_0$ is reached, and once asymptotically after $V_{max} \equiv V_0$ is reached.
\\\\
It follows that (\ref{eqn5d}) can be consistently interpreted in this special case if $M_{flat}$ is understood to be the contained luminous mass up to the \emph{first} occurrence of $V_{flat}$ on the rotation curve. It is this value of $M_{flat}$ which satisfies the BTFR in the theoretical development of \S\ref{BTFR}. It is clear that $M_{flat}{\rm(theory)} < M_{flat}{\rm(photometry)}$ for this case but, in practice, we find that the corrections required to make  the definition of $M_{flat}$(photometry) consistent  are too small to make any noticeable difference to the analysis of  \S\ref{Mflat}.
\subsection{Whole-disk scaling relation for mass}
We use the quantitative statement of Freeman's Law, (\ref{eqn4e}), and the definition of $M_{flat}$ given at (\ref{eqn5d}) to derive,  in appendix \S\ref{ScalingLaw}, the scaling laws used in \S\ref{RAR} to demonstrate the RAR in conservative MOND, and given below:
\begin{eqnarray}
M(R) &=& M_0  \left( \frac{R}{R_0}\right)^2\left( \frac{V_0}{V_{rot}(R)}\right)^2,~~~ R  \leq R_0, \nonumber \\
\label{RAR3} \\
M(R) &=& M_0 \left( \frac{V_{rot}(R)}{V_0}\right)^2,~~~ R > R_0,~~V_{rot}(R) \leq V_{flat}.
\nonumber
\end{eqnarray}
The advantage of using these is simply that they require no explicit reference to $\mathcal{M}_g(R)$, the mass distribution on $R\leq R_0$. In fact, $\mathcal{M}_g(R)$ enters these scaling laws through the calculated theoretical values of $(M_0,V_0)$.
\section{The conservative MOND modelling process} \label{SA} 
Classical MOND uses   photometrically determined mass-modelling within galaxy disks, in conjunction with photometrically determined distance scales, to predict the details of the associated rotation curves. Conservative MOND is applied in the reverse way: we use dynamical modelling of the rotation curves to determine the absolute distance scales, the  characteristic mass parameter $M_0$ and the total radial distribution of disk mass up to $M_{flat}$  for each disk, which we then compare against the photometric estimations (SPARC photometry) of the same quantities. 
\subsection{Choice of the mass function $\mathcal{M}_g(R)$} \label{MassF}
For simplicity, we are choosing to model mass orbiting in a disk galaxy as mass orbiting in the equatorial plane of a spherical space in which the mass contained with radius $R$ is given by $\mathcal{M}_g(R)$. So, the question is: \emph{How do we choose this mass function?}
\\\\
We know, from (\ref{eqn2}), that
\[
V_{rot}(R) = V_{flat} \,\left(\frac{ \Sigma_F }{   \Sigma_R } \right)^{1/2} 
\]
where $\Sigma_R$ is the mass surface density on the sphere radius $R$ centred on the galactic object whilst $\Sigma_F$ is the characteristic mass surface density of the fractal $D\approx 2$ external environment. Thus, in our simplistic spherical model, at any given radius $V_{rot}(R)$ is independent of the detailed structure of the mass distribution contained within that radius.
\\\\
To illustrate the point, if we choose the simple model
\begin{equation}
\mathcal{M}_g(R) \equiv M_0 \left(\frac{R}{R_0} \right)^{3/2},~~~~R \leq R_0, \label{MF2}
\end{equation}
then  $\Sigma_R \sim R^{-1/2}$. This behaviour is consistent with the sphere containing, for example, a constant thickness disk with volume density behaving as $\rho \sim R^{-1/2}$, or a whole class of similar configurations. So, whatever model we choose, that model is best considered as simply a constraint on the possibilities for a disk galaxy contained within a spherical volume. 
\\\\
In practice, we find that the simple model above, used across the whole SPARC sample of non-bulgy objects, gives a reasonable broad-bush picture for present purposes.
\subsection{The algorithmic details}\label{Details}
The mass function $\mathcal{M}_g(R)$, to be used across the whole SPARC sample of non-bulgy objects, is defined as the simple model (\ref{MF2}).
Equation (\ref{Frac6}) can be written as:
\begin{equation}
V_0^2 =  V_{flat}^2 \left( \frac{ a_0  R_0^2}{G M_0}\right).  \label{Frac9}
\end{equation}
Taking note of some essential computational details described in \S\ref{MassModels}, the scaling relations of (\ref{eqn2})  can now be applied as follows:
\begin{enumerate}
\item  The modelling algorithm treats the three parameters $( R_0, M_0,  V_{flat})$ as independent, all to  be varied in order to optimize the fit of $V_{rot}(R)$ given at (\ref{eqn2}) to the SPARC rotation curve in the disk concerned. An automatic code, based on the Nelder-Mead method (robust on noisy data), is used for this process;
\item Compute $V_0$ from (\ref{Frac9}). The full set of characteristic parameters $(R_0, V_0, M_0, V_{flat})$ computed according to the photometric distance scalings implicit to SPARC for the given disk are then available;
\item At this stage, all computations have been done assuming the distance scales implicit to the SPARC sample so that, generally speaking, the requirements of the MOND acceleration condition, $V_0^2/R_0 = a_0$, are not met. In fact, we routinely find
\[
M_0({\rm theory}) >> M_0({\rm photometry})
\] 
so that, as with conventional theory, a \lq{missing mass}' problem has emerged; 
\item But conservative MOND requires the MOND acceleration condition to be satisfied. So, normalize the distance scale for each disk according to $R\rightarrow  K R$, where $K$ is chosen to ensure that $V_0^2/R_0 = a_0$ is exactly satisfied; 
\item When $R$ is normalized in this way, then (\ref{Frac9}) shows that the calculated values of $M_0$ must be normalized according $M_0\rightarrow  K^2 M_0$ in order to ensure that the velocities remain invariant;
\item A comparison with SPARC photometry now shows:
\[
M_0({\rm theory}) = M_0({\rm photometry})
\] 
 at the level of statistical certainty over the whole of the SPARC sample. There is no mass discrepancy;
\item We can now compute the radial distribution of mass over $R>0$  up to $V_{flat}$ using the scaling relations (\ref{RAR3}) to find:
 \[
 M(R)({\rm theory}) = M(R)({\rm photometry}),~~~R > 0,~~~ V_{rot}(R) \leq V_{flat}
 \] 
at the level of statistical certainty over the whole of the SPARC sample. There is no mass discrepancy;
\end{enumerate} 
 As we shall see in \S\ref{SPARCmass1} and \S\ref{SPARCmass2}, the statistical power of these results represents 
 compelling evidence in support of the mass rescaling process and, by inference, the distance rescaling process.
\subsection{Dealing with noisy data: The Least-Area regression technique} \label{LAR}
In general, the foregoing requires that we ask whether the relationship 
\begin{equation}
Mass({\rm theory}) \approx Mass(\rm{photometric}) \label{Mass1}
\end{equation}
is supported on the data.
As we shall see, the correlation between these two quantities, after rescaling using the MOND acceleration condition (but not before), is statistically extremely powerful.
\\\\
However, notwithstanding this powerful correlation,  both quantities are very noisily determined, and so the standard tool of least-squares linear regression is particularly poorly suited for the task of determining any quantitative relationship between them, since this standard tool assumes the predictor to be entirely free of error. In consequence, any quantitative relationship we deduce between predictor and response  using least-squares linear regression is always strongly dependent on the choice of the predictor, and therefore cannot be relied upon in the present context.
\\\\
We eliminate this problem by developing a method of linear regression based on \emph{least areas}, which treats predictor data and response data in an entirely symmetric fashion. Consequently, the quantity being minimised is independent of how the predictor/response pair is chosen for the regression. The result is a linear model which can be algebraically inverted to give the exact linear model which would also arise from regressing on the interchanged predictor/response pair. So, any inference drawn about the relationship between the predictor and response is independent of how the predictor/response pair is chosen. 
The details are given in \S\ref{LeastAreas}.
\section{Theory against observation: preliminary comments} \label{Prelim}
The SPARC sample compiled by \citet{McGaugh2015} consists of 175 nearby galaxies with modern surface photometry at $3.6\,\mu m$ and high quality rotation curves. The sample has been constructed to span very wide ranges in surface brightness, luminosity, rotation velocity and Hubble type, thereby forming a representative sample of galaxies in the nearby Universe. To date, the SPARC sample is the largest collection of galaxies with both high-quality rotation curves and NIR surface photometry.
\\\\
We consider only the SPARC galaxies which have quality flag Q = 1 or 2, which gives a total sample of 160 objects out of a total of 175 objects. For simplicity, and ease of making the main points, we then select only those objects which appear to have no measurable bulge component (that is, are explicitly stated to have no measurable bulge component in the database) and have inclination $> 30^o$, giving a final sample of 118 objects.
\\\\
Given that a fixed $MLR = 2.0$ is applied to the photometry across the whole SPARC sample, then the only parameters varied in the algorithm of \S\ref{Details} to optimize RC fits are the three characteristic disk parameters $(R_0, M_0, V_{flat})$ from which, using (\ref{Frac9}),  we obtain the full set of characteristic disk parameters $(R_0, V_0, M_0, V_{flat})$. 
\\\\
The critical acceleration boundary, $R=R_0$, separates the galactic interior from the exterior IGM medium. For this reason, it is natural to present the results for $R\leq R_0$ and $R > R_0$ separately, which we do in \S\ref{SPARCmass1} and \S\ref{SPARCmass2} respectively.
\\\\
These calculations provide quantitative estimations of the luminous mass distribution across the whole disk of each object in the sample, derived purely from disk dynamics.
\subsection{SPARC LSBs}\label{LSBs}
Before considering the results against SPARC photometry  in detail, it is germane to recognize that, according to the notes in the source papers used by \citet{McGaugh2015} to compile the SPARC sample, that sample contains at least 25 LSB objects, listed as:  F563-1, F568-1, F568-3, F568-V1,  F571-V1, F571-8, F574-1, F583-1, F583-4, NGC3917, NGC4010,  UGC00128, UGC01230, UGC05005,  UGC5750,  UGC05999, UGC06399, UGC06446, UGC06667,  UGC06818,  UGC06917,  UGC06923, UGC06930, UGC06983, UGC07089.
\\\\ 
Since such objects are defined as ones in which the entire disk is in the MOND weak-gravity regime then, for these objects, $R_0=0$,  meaning that, at face value, the MOND acceleration condition cannot be used to scale these objects. For this reason, in the figures \ref{fig:SPARCMASSvsTheoryMass} \& \ref{fig:SSM-4} of \S\ref{SPARCmass1} and figures \ref{fig:SSM-2} \& \ref{fig:SSM-5} of \S\ref{SPARCmass2} we identify these LSB objects as filled red circles so that their behaviour under the analysis can be observed. 
\\\\
Against expectation, and as is absolutely clear from the figures, we find that the LSB objects behave exactly as the non-LSB objects under the MOND acceleration condition.   In practice, this means that, in fact, the objects classified as LSBs in the SPARC sample all have an objectively detected critical boundary $R=R_0>0$ in their disks which has allowed the automatic \& successful application of the MOND acceleration condition to the objects concerned. 
\\\\
In other words, none of the listed objects is an LSB in the sense of the standard definition; the conflict is consistent with the idea that there is an unrecognized mechanism causing the standard photometric distances for these objects to exaggerate the actual distances, giving rise to the appearance that the disks concerned are all in the deep MOND regime. 
\section{Theory vs SPARC photometry: $R\leq R_0$}\label{SPARCmass1}
All error bars given in the least-area linear regressions are set at the conventional two standard deviations, determined by a bootstrapping process.
\subsection{$M_0 \equiv \mathcal{M}_g(R_0)$\,: the total mass inside the critical radius, $R = R_0$}\label{M0}
The parameter $M_0$ represents the predicted total mass inside the critical radius $R = R_0$ for the galaxy concerned. To obtain the SPARC photometric estimate of the same quantity we simply adopted a global mass-to-light ratio in the disks of $MLR = 2.0$, and then integrated the disk photometry over all $R \leq R_0$ (taking care to use original SPARC scalings) to obtain the photometric estimates of $M_0$. Although this represents a very crude way of representing the contributions of the various components of mass within $R = R_0$, it is very effective for current purposes. The results  for the whole sample of 118 objects are displayed in the two panels of Figure\,\ref{fig:SPARCMASSvsTheoryMass} and discussed below. 
\subsubsection{Fig \ref{fig:SPARCMASSvsTheoryMass} Upper: Distance scales from conventional photometry} \label{2U}
\begin{figure}[H]
	\centering
	\includegraphics[width=0.7\linewidth]{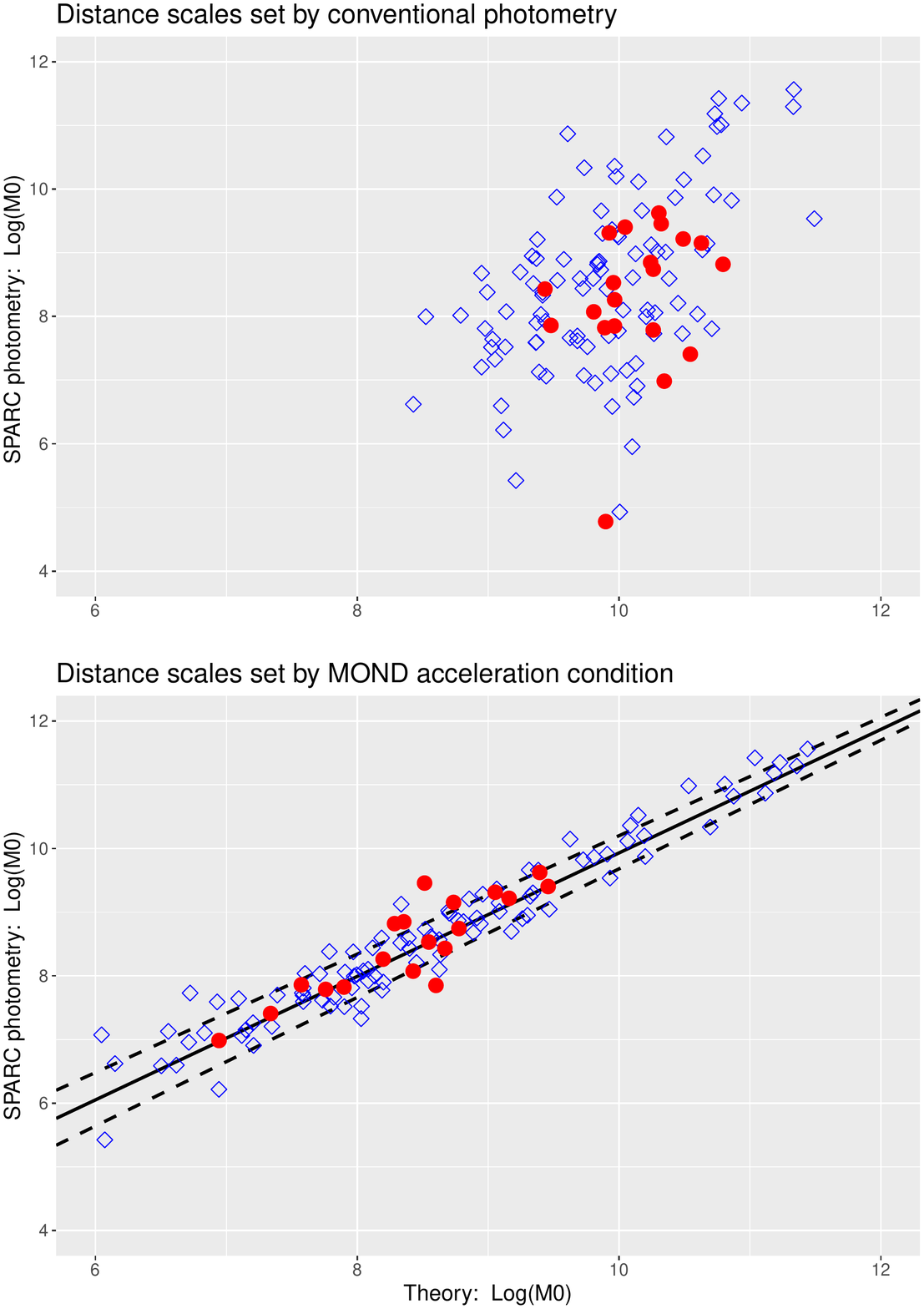}
	\caption{ Red filled circles = putative LSB objects. Open diamonds = everything else. Upper figure:  The $\log{M_0}$(theory) values have been computed on the conventional photometric scalings implicit to the SPARC sample. Lower figure: The upper panel $\log{M_0}$(theory) values have been rescaled according to the requirements of the MOND acceleration condition.}
	\label{fig:SPARCMASSvsTheoryMass}
\end{figure}

Here, putative LSB objects are identified as red filled circles, and $M_0$(theory) is computed by the algorithm of \S\ref{Details} using the photometric distance scalings implicit to the SPARC sample.
The  scatter plot for this case makes it clear that for the most massive objects, say $10 \leq M \leq 12$, there is a rough qualitative agreement between theory and photometry. However, at the low photometric end, we see that $M{\rm(theory)}> M{\rm(photometry)}$ by up to two orders of magnitude. That is, a \lq{missing mass}' problem of similar proportions to that recognized in classical gravitation theory has emerged.
\subsubsection{Fig \ref{fig:SPARCMASSvsTheoryMass} Lower:  Distance scales from MOND acceleration condition} \label{2L}
Here, putative LSB objects are again identified as red filled circles and $M_0$(theory) computed above has been rescaled according to the requirements of the MOND acceleration condition (\ref{eqn3NN}), specified in \S\ref{Details}.
The  scatter plot for this case makes it clear that this rescaling process has the effect of mapping every data point almost perfectly onto the $\log M_0$(theory) $=\log M_0$(photometry) line. That is, there is now an almost statistically perfect correspondence between the two mass quantities and a least-area linear regression gives:
\[
\log{M_0}({\rm photometry}) \approx \left(0.99 \pm 0.07\right) \log{M_0}({\rm theory}) + (0.19 \pm 0.62).
\]
In other words, the imposition of the conditions required by the MOND acceleration condition has the effect of imposing an almost perfect correspondence  $M_0{\rm(theory)} \approx M_0{\rm(photometry)}$, so that the \lq{missing mass}' problem of the upper panel has disappeared completely. Note that the behaviour of the  LSB objects cannot be distinguished from that of the non-LSB objects. We have discussed this detail in \S\ref{LSBs}.
\subsection{The radial distribution of mass on the interior: $R < R_0$} \label{Interior}
Using a fixed MLR=2.0, we have integrated the photometry for each disk in the SPARC sample to obtain estimates of $M(R_i)(\rm{photometry})$ contained within radius $R_i$, for each radial coordinate $R_i < R_0$ on the measured disk; this gives a total of 374 individual estimated mass values over the whole non-bulgy sample. In the following, we compare these against theoretical estimates of the same quantities given by the conservative MOND algorithmic process defined in \S\ref{Details}.
\subsubsection{Fig \ref{fig:SSM-4}  Upper: Distance scales from conventional photometry}
\begin{figure}[H]
	\centering
	\includegraphics[width=0.7\linewidth]{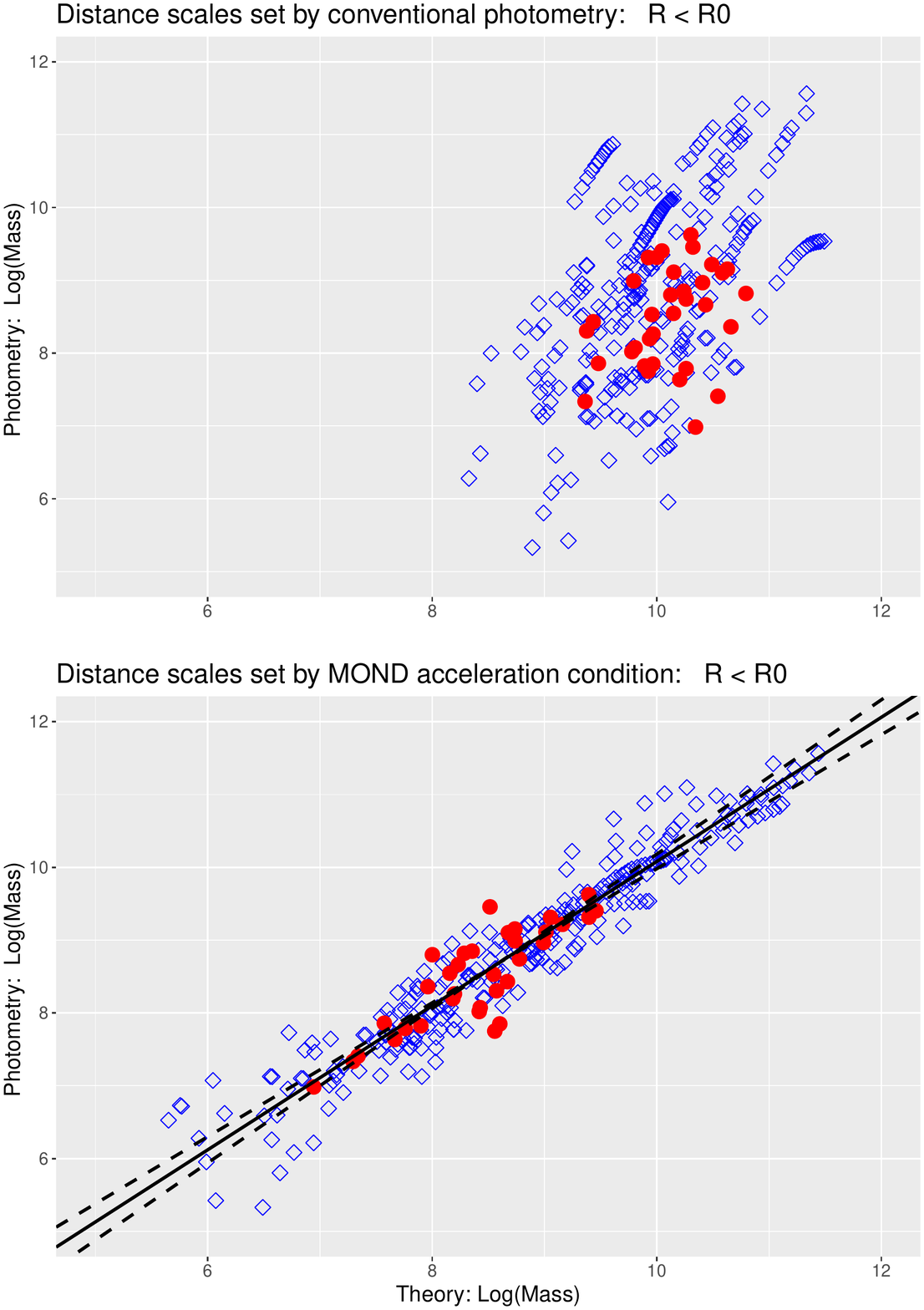}
	\caption{ Red filled circles = putative LSB objects. Open diamonds = everything else. Upper figure:  The  $\log{M(R)}$(theory) values for all disk data points satisfying  $R \leq R_0$ (308 individual points in total) have been computed using the conventional photometric distance scalings implicit to the SPARC sample.  Lower figure: The 308 $\log{M(R)}$(theory) values of the upper panel  have been rescaled according to the requirements of the MOND acceleration condition. }
	\label{fig:SSM-4}
\end{figure}

Ditto comments of \S\ref{2U}.
\subsubsection{Fig \ref{fig:SSM-4} Lower: Distance scales from MOND acceleration condition}
Ditto comments of \S\ref{2L}. There is now an almost statistically perfect correspondence between the two quantities and a least-area linear regression gives for all objects:
\[
\log{M}({\rm photometry}) \approx \left(1.02 \pm 0.05\right) \log{M}({\rm theory}) - (0.07 \pm 0.50),
\]
and for LSBs only:
\[
\log{M}({\rm photometry}) \approx \left(0.99 \pm 0.07\right) \log{M}({\rm theory}) - (0.18 \pm 0.60),
\]
som that $\log M(\rm{theory}) \approx \log M(\rm{photometry})$ is confirmed for all the mass measurements inside the critical radius, $R \leq R_0$.
In other words, the imposition of the conditions required by the MOND acceleration condition has the effect of making the \lq{missing mass}' of the upper panel disappear completely. 
\section{Theory vs SPARC photometry: $R > R_0$}\label{SPARCmass2}
There is a fundamental distinction between what happens on $R\leq R_0$ and what happens on $R>R_0$, which must be carefully accounted for.
\\\\
The process of deriving conservative MOND outlined in \S\ref{OverView} required the galactic object, represented by  $\mathcal{M}_g(R\leq R_0)$, to be placed into  an unbounded $D=2$ fractal material environment as follows: 
\begin{eqnarray}
\mathcal{M}(R) &=& \mathcal{M}_g (R),~~~ R \leq R_0; \nonumber \\
\mathcal{M}(R) &=&  \mathcal{M}_g(R_0) + 4 \pi \Sigma_F \left(R^2 - R^2_0 \right),~~~ R > R_0 \nonumber
\end{eqnarray}
where $\mathcal{M}(R)$ represents the radial distribution of total mass on $0<R<\infty$. 
This theoretical structure is interpreted as a representation of a galactic object sitting in a hypothetical $D \approx 2 $ quasi-fractal grey dust IGM which, for the reasons discussed in \S\ref{Distances1}, would be extremely difficult, if not impossible, to detect.
\\\\
The distinction between events on $R\leq R_0$ and events on $R>R_0$ is now obvious: In \S\ref{SPARCmass1} it was shown that all of the core object mass represented by $\mathcal{M}_g(R\leq R_0)$ was photometrically accounted for. But we know that on $R>R_0$, the total of all photometrically detectable mass, $M_{flat}$(photometry) say, is finite and bounded, whereas the total mass available on $R>R_0$ is unbounded.
\\\\
So, given that the scaling relationships on $R>R_0$ actually do account for  the photometrically detectable mass only (and not the generally available mass) we must ask \emph{what is going on?}   
\\\\
The only consistent interpretation of these circumstances would appear to be that the photometrically detectable  material in the region $R>R_0$ is detectable because it is excited to radiate above the background by the gravitational action of the core object, $\mathcal{M}_g(R \leq R_0)$. In old-fashioned terms, the core works the disk, thereby heating it. The  scaling relations then simply describe the effect of that action upon the disk and hence determine the mass of excited material and its radial distribution.
\subsection{$M_{flat}$\,: Theory compared to SPARC photometry}\label{Mflat}
The quantity $M_{flat}$ is conventionally defined as the mass contained within the disk out to $V_{rot} \approx V_{flat}$, and is estimated by integrating disk-photometry; consequently, by definition, $M_{flat}$ quantifies the detectable \emph{luminous} mass. However, in \S\ref{BTFR}, we found that the key to deriving the explicit BTFR, given at (\ref{eqn5c}) as
\begin{equation}
V_{flat}^4 = a_0\, G\, M_{flat}{\rm(theory)}, \label{eqn5b} 
\end{equation}
was the definition of (\ref{eqn5d}) that: 
\begin{equation}
M_{flat}{\rm(theory)} \equiv M_0\,\left(\frac{V_{flat}}{V_0} \right)^2  \label{eqn5e}
\end{equation}
which was key (appendix \S\ref{ScalingLaw}) to extending the scaling relationship (\ref{RAR3}) to the region $R > R_0$.
So the initial question to be answered is whether this definition  tracks $M_{flat}$(photometric) determined, for the SPARC sample, by  \citet{McGaugh2015}? 
\\\\
For this exercise,  we are testing $M_{flat}$(theory) directly against the specific $M_{flat}$(photometry) values provided by authors of the SPARC sample, and so take no account at all of the few special cases for which $V_0> V_{flat}$, discussed in \S\ref{SpecialCase-1}.
\subsubsection{The study of \citet{McGaugh2015}}
The work of \citet{McGaugh2015} was motivated by the idea that, within $\Lambda$CDM cosmology, the BTFR can only emerge from a complex process of galaxy formation, and is hence expected to be associated with significant intrinsic scatter. In short, the degree to which intrinsic scatter is present within the BTFR provides a key test for $\Lambda$CDM cosmology. The very high-quality of the SPARC sample provided an ideal opportunity to investigate BTFR scatter in a sample of substantial size. Subsequently, the authors were able to show that the SPARC sample is highly constrained by the BTFR showing far less scatter that expected from the $\Lambda$CDM model.
\\\\
But their results, in demonstrating a very tight fit of the empirical BTFR to SPARC data, also provide a test of  degenerate case conservative MOND: specifically, to test whether or not $M_{flat}$(theory)  defined according to the scaling relation (\ref{eqn5e}) follows the photometric determinations of the same quantity given by \citet{McGaugh2015} over a sample of 118 SPARC objects.
\\\\
Our basic sample consists of the 118 SPARC non-bulgy objects with quality flag Q = 1,2 and inclination $> 30^o$. By contrast, the relevant sample of  \citet{McGaugh2015} consists of the 118 SPARC objects for which good estimations of $V_{flat}$ (using the standard photometric algorithm) were possible. Since some of these are bulgy objects (excluded from our sample), this gave a final sample of 83 SPARC objects of which to test $M_{flat}$(theory), given by (\ref{eqn5e}), against $M_{flat}$(photometry) given by  \citet{McGaugh2015}. 
\\\\
The results are displayed in the two panels of Figure\,\ref{fig:SSM-2} and discussed below.
\subsubsection{Fig \ref{fig:SSM-2} Upper: Distance scales from conventional photometry}
\begin{figure}[H]
	\centering
	\includegraphics[width=0.65\linewidth]{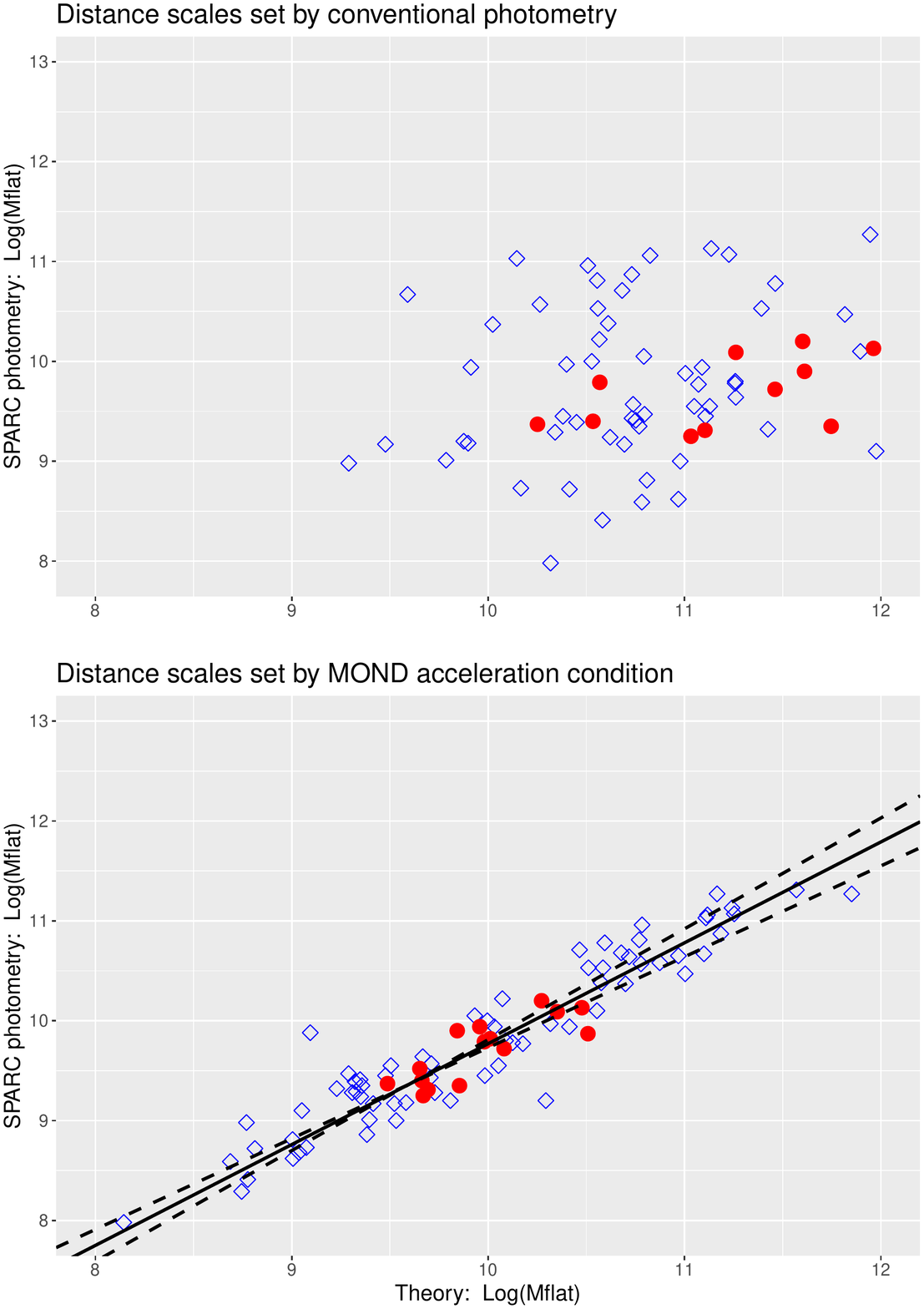}
	\caption{ Red filled circles = putative LSB objects. Open diamonds = everything else. Upper figure:  Here, the $\log{M_{flat}}$(theory) values have been computed using (\ref{eqn5e}) on the conventional photometric scalings implicit to the SPARC sample.  Lower figure: Here, the upper panel $\log{M_{flat}}$(theory) values have been rescaled according to the requirements of the MOND acceleration condition.  No outliers have been removed.}
	\label{fig:SSM-2}
\end{figure}

Ditto comments of \S\ref{2U}. 
\subsubsection{Fig \ref{fig:SSM-2} Lower:  Distance scales from MOND acceleration condition}
Ditto comments of \S\ref{2L}. There is now an almost statistically perfect correspondence between the two quantities and a least-area linear regression gives:
\begin{equation}
\log M_{flat}({\rm photometry}) =   \left(1.01 \pm 0.10 \right) \log M_{flat}({\rm theory}) -  \left(0.33 \pm 0.96 \right).  \label{eqn6d}
\end{equation}
Again, the imposition of the conditions required by the MOND acceleration condition has the effect of imposing an almost perfect correspondence  $M_{flat}{\rm(theory)} \approx M_{flat}{\rm(photometry)}$, so that for all practical purposes, $M_{flat}$(theory) determined by (\ref{eqn5e}) tracks the \citet{McGaugh2015} photometric determinations of $M_{flat}$ with high statistical fidelity. The \lq{missing mass}' problem of the upper panel has disappeared completely.
\\\\
Finally, for completeness, a least-area regression between $\log M_{flat}$(photometry) and $\log V_{flat}$(theory) gives:
\begin{equation}
\log M_{flat}({\rm photometry}) =   \left(4.05 \pm 0.38 \right) \log V_{flat}({\rm theory}) +  \left(1.49 \pm 0.80 \right)  \label{eqn6dd}
\end{equation}
so that the BTFR is again confirmed on the SPARC data. A comparison between (\ref{eqn6dd}) and the corresponding regressions in 
\citet{McGaugh2015} shows that they are statistically identical, including the zero points. The only difference is that we are using $V_{flat}$(theory) rather than $V_{flat}$(SPARC).
\subsection{The radial distribution of luminous mass on the exterior: $R > R_0$} \label{Exterior}
We use the scaling relation (\ref{RAR3}) on $R > R_0$ so that
\begin{equation}
M_{Lum}(R) =  M_0 \left( \frac{V_{rot}(R)}{V_0} \right)^2,~~~R >R_0,~~~V_{rot}(R) \leq V_{flat}.~~~ \label{eqn7}
\end{equation}
In order to test the scaling law (\ref{eqn7}), we have integrated the photometry for each disk in the SPARC sample to obtain estimates of $M_{Lum}(R_i)(\rm{photometry})$ contained within radius $R_i$, for each radial coordinate $R_i > R_0$ on the measured disk; this gives a total of 1608 individual estimated mass values over the whole non-bulgy sample. The results for this test of (\ref{eqn7})  are shown in figure \ref{fig:SSM-5}.
\subsubsection{Fig \ref{fig:SSM-5} Upper: Distance scales from conventional photometry}
\begin{figure}[H]
	\centering
	\includegraphics[width=0.7\linewidth]{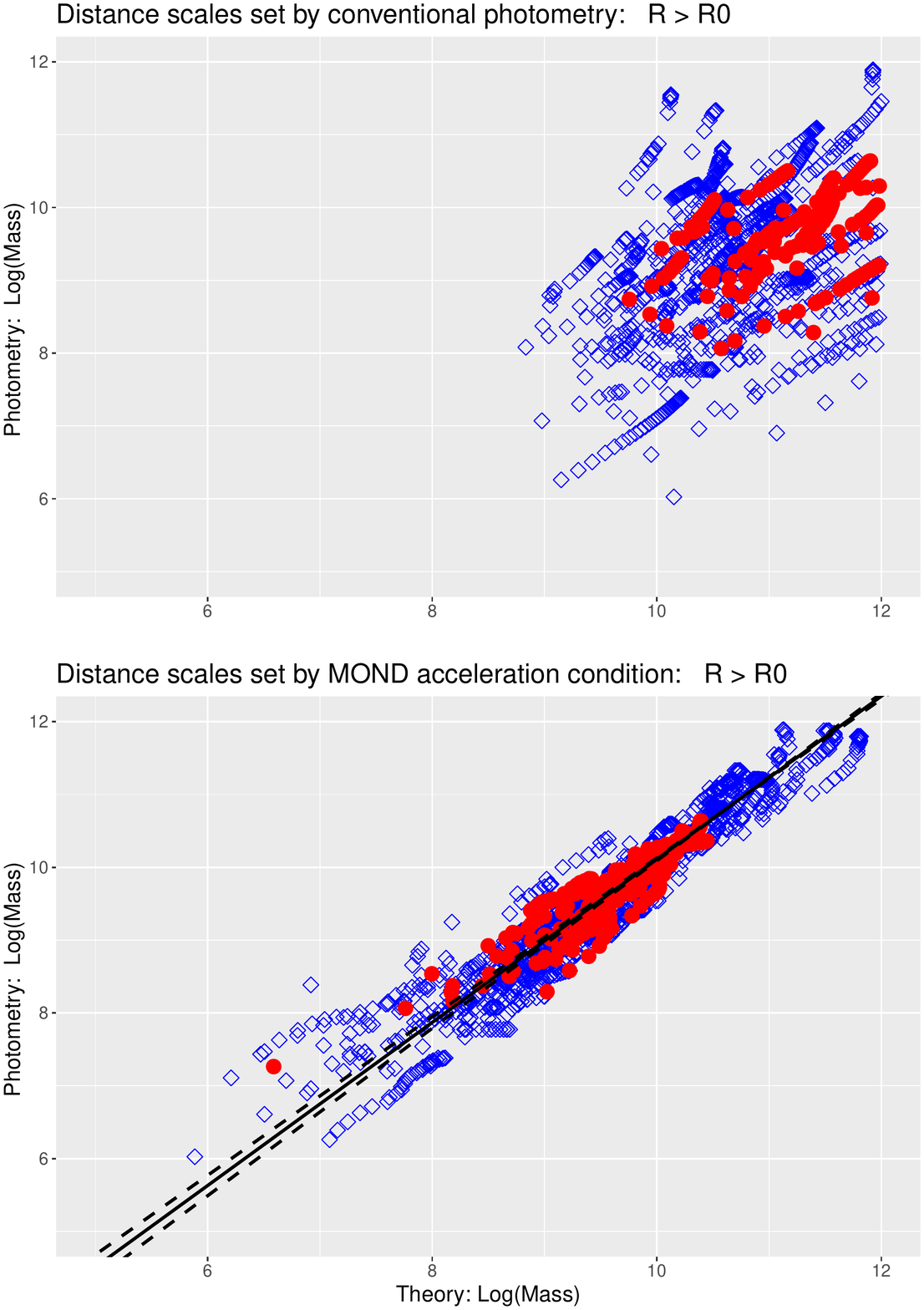}
	\caption{ Red filled circles = putative LSB objects. Open diamonds = everything else. Upper figure:  Here, the  $\log{M}(R)$(theory) values for all disk data points satisfying  $R > R_0$ (1608 individual points in total) have been computed using the conventional photometric distance scalings implicit to the SPARC sample.  Lower figure: Here, the 1608 $\log{M(R)}$(theory) values of the upper panel  have been rescaled according to the requirements of the MOND acceleration condition.}
	\label{fig:SSM-5}
\end{figure}

Ditto comments of \S\ref{2U}.
\subsubsection{Fig \ref{fig:SSM-5} Lower: Distance scales from MOND acceleration condition}
Ditto comments of \S\ref{2L}.
There is now an almost statistically perfect correspondence between the two quantities and a least-area linear regression gives for all objects:
\[
\log{M}({\rm photometry}) \approx \left(1.12 \pm 0.03\right) \log{M}({\rm theory}) - (1.09 \pm 0.32), 
\]
and for LSBs only:
\[
\log{M}({\rm photometry}) \approx \left(1.05 \pm 0.11\right) \log{M}({\rm theory}) - (0.44 \pm 1.07), 
\]
so that $\log M(\rm{theory}) \approx \log M(\rm{photometry})$ is confirmed for all the mass measurements outside the critical radius, $R > R_0$.
\\\\
We can unambiguously conclude that the scaling relation (\ref{eqn7}) is validated on the SPARC sample.
\section{Qualitative behaviour of RCs on $R>R_0$} \label{RCbehaviour-1}
It is easily shown from (\ref{eqn2}c) that, for $R>R_0$: 
\[
\frac{d V_{rot}}{dR} \sim \frac{k^2}{R^3} \left( \Sigma_0 - \Sigma_F \right),~~~k^2>0,~~~R>R_0
\]
where $\Sigma_0 \equiv \mathcal{M}_g(R_0)/(4\pi R_0^2)$.
As a direct consequence, we can then deduce:
\begin{itemize}
	\item if $\Sigma_0 > \Sigma_F$, then rotation curves on $R\leq R_0$ reach $(R_0,V_0)$ and then rise smoothly with an asymptotic approach to a constant rotation velocity $V_{flat} > V_0$; 
	\item if $\Sigma_0 = \Sigma_F$, then rotation curves on $R \leq R_0$ reach $(R_0,V_0)$ and then change abruptly to flatness with constant rotation velocity $V_{flat} = V_0$;
	\item if $\Sigma_0 < \Sigma_F$, then rotation curves on $R\leq R_0$ reach $(R_0,V_0)$ and then fall smoothly with an asymptotic approach to a constant velocity $V_{flat}< V_0$. 
\end{itemize}
\section{Summary and conclusions}
 \subsection{Summary}
The foregoing for all objects can be entirely summarized by reviewing the case of LSB galaxies.
\\\\ 
MOND and conservative MOND have in common Milgrom's fundamental idea of a critical acceleration, $a_0 \approx 1.2 \times 10^{-10} m/sec^2$.  For classical MOND, the radius at which the MOND acceleration condition, $V_0^2/R_0=a_0$, is observed to be satisfied defines the position at which one gravitational law gives way to another. By contrast, for conservative MOND, the critical boundary, $R=R_0$, is determined by the algorithm of \S\ref{SA} which is independent of calculating apparent centripetal accelerations. It follows immediately that the MOND acceleration condition provides the means for normalizing distance scales which is completely independent of standard candles and the photometric method:  we simply adjust the distance to the galaxy concerned (and hence all associated linear scales)  until the centripetal acceleration (used as a proxy for gravitational acceleration) at the critical boundary satisfies $V^2/R \approx a_0$.
\\\\
Now consider the case of low surface brightness disk galaxies: LSB disks are defined as disks for which $V^2/R < a_0$ everywhere in the disk, and it is considered to be one of MOND's great successes that such apparently weakly gravitating objects were predicted to exist before they were first observed in the early 1990s. It follows that within LSB disks, there is no critical boundary to detect, and therefore the MOND acceleration condition cannot be used to set a distance scale for them.
\\\\
But  consider the SPARC sample: according to the notes in the source papers used by \citet{McGaugh2015} to compile that sample, it contains at least 25 LSB objects, listed in \S\ref{LSBs}. But figures \ref{fig:SPARCMASSvsTheoryMass}, 	\ref{fig:SSM-4}, \ref{fig:SSM-2} \&  \ref{fig:SSM-5} demonstrate beyond all doubt that these putative LSB objects behave exactly as do all the other objects in the SPARC sample under the MOND acceleration condition. In other words, as a matter of objective fact, they must necessarily all possess a critical boundary $R=R_0$ which has been successfully located by the algorithm of \S\ref{SA}, and then automatically acted upon via the requirements of the acceleration condition, with the final result of mapping $Mass$(theory) exactly (in a statistical sense) onto $Mass$(photometric).
\\\\
There is only one way this objective fact can be made consistent with the appearance that $V^2/R < a_0$ over the resolvable disk for all the objects classified as LSBs: there must be some unrecognized mechanism which systematically exaggerates distance scales, and hence the linear scales of these objects, when these are determined photometrically, giving rise to the \emph{appearance} that $V^2/R < a_0$ everywhere over the resolvable disk.
\subsection{Conclusions}
Figures  \ref{fig:SPARCMASSvsTheoryMass}, \ref{fig:SSM-4}, \ref{fig:SSM-2}, \& \ref{fig:SSM-5} demonstrate beyond all doubt that:
\begin{itemize}
	\item Conservative MOND, used in conjunction with the photometric distance scales implicit to the SPARC sample, returns mass distributions that imply a missing mass problem of similar magnitude to canonical theory;
	\item Normalizing the distance scale, $R \rightarrow K R$, to ensure the condition $V_0^2/R_0 = a_0$ at the critical boundary (determined by the algorithm of \S\ref{SA}) requires that masses are normalized according to $M \rightarrow K^2 M$ in order to keep velocities invariant;
	\item This mass normalization maps $M$(theoretical)  onto $M$(photometric) with perfect statistical fidelity. The missing mass problem goes away;
	\item The implication of all this is that photometric distance scales systematically inflate true distance scales, whilst normalizing according to the MOND acceleration condition applied at the critical boundary returns a true record;
	\item A simple explanation for the failure of the photometric method for estimating distance scales is that the $D \approx 2$ quasi-fractal IGM (which is integral to conservative MOND)  consists of \lq{grey dust}' which causes extinction without reddening - for example, see \citet{AA1999} or  \citet{PSC2006}. See discussion of \S\ref{Distances2};
	\item For the reasons discussed in  \S\ref{Distances1} and \S\ref{Distances2}, such a grey-dust distribution would be almost impossible to detect by any direct means.
\end{itemize}
The utility of degenerate case conservative MOND for rotationally supported systems is established. Given data of similar quality to that of the SPARC sample, general case conservative MOND would be implemented in a similar way, using a modified version of the \S\ref{SA} algorithm. For the moment, data of the required quality is not available.
\appendix
\section{Leibniz-Mach cosmology: Overview} \label{LMC}
The Leibniz-Mach cosmology appeared in the mainstream literature in a primitive form as \citet{Roscoe2002}. This early form is mathematically complete, but only partially interpreted around the meaning of clocks in the equilibrium state. The fully interpreted form is given as online supplementary material here, and exists in the archive as \citet{Roscoe2020}.
The following provides an outline of the cosmology.
\subsection{Briefly, Leibniz on space} 
In the
debate of Clarke-Leibniz (1715$\sim$1716) (\citet{Alexander1984}) concerning the nature of space, and in which
Clarke argued for Newton's conception of an absolute space, Leibniz made three arguments
of which the second was:
\begin{quote}
	\emph{Motion and position are real and detectable only in relation
		to other objects ... therefore empty space, a void, and so space itself
		is an unnecessary hypothesis.} 
\end{quote}
Of Leibniz's three arguments, this latter was the only one to which
Clarke had a good objection - essentially that \emph{accelerated motion}
can be perceived without reference to
external bodies and is therefore, he argued, necessarily perceived
with respect to the absolute space of Newton.
Given Clarke's objection, which was pertinent, and whatever Leibniz's actual intention, it is clear that the world implicit
to his argument above is a world in which motion is everywhere uniform and unaccelerated - otherwise Clarke's objection must stand. 
So, Leibniz's  implied equilibrium state  leads us to consider whether or not such a thing can actually exist, and hence we are led to  that which we nominate as \emph{The Leibniz Question:}
\begin{quote}
	\emph{Is it possible to conceive a non-trivial global mass distribution, isotropic about every spatial origin,  which is such that motions are everywhere uniform and unaccelerated  and, if so, what are the properties of this distribution?} 
\end{quote}
The Leibniz Question was a primary driver in the development of the Leibniz-Mach cosmology and, as we shall see, the answer to the Leibniz Question is 
\begin{quote}
\emph{Yes, and the distribution of material is necessarily fractal, $D=2$. }
\end{quote}
\subsection*{Briefly, Leibniz \& Mach on time}
\citet{Mach1960} was equally clear in expressing his views about the nature of
time which are, in effect, very similar to those expressed by Leibniz.  They each viewed \emph{time} (specifically Newton's \emph{absolute
	time}) as a meaningless abstraction. All that we can ever do, Mach argued, is to measure \emph{change} within one system against
\emph{change} in a second system which has been defined as the standard
(eg it takes half of one complete rotation of the earth about its
own axis to walk thirty miles).
So, on this Machian view, the `clock' used to quantify the passage of time for a physical system A is simply an independent physical system B which has been arbitrarily defined as the standard clock.
\\\\
It follows from the foregoing considerations that any definition of \emph{time} based upon the Machian argument must be based upon the ideas of: 
\begin{itemize}
	\item every physical system having its own subjective and private internal time-keeping, any one of which can be chosen as the \emph{standard clock} against which all the other systems reckon the passage of time;
	\item in practice, one such system being chosen as the standard clock, the choice being merely one of convention usually involving a natural cyclic process.
\end{itemize}
This ultimately leads to the qualitative result that:
\begin{quote}
	\emph{Every particle and every system of particles is no more than an angular momentum conserving clock. So, in particular, every disk galaxy is nothing more than a giant  angular momentum conserving clock.}
\end{quote}
\subsection{Synthesis of quantitative model} \label{SQM}
In modern terms, the net conclusions arising from the detailed analysis of how clocks and rods are to be defined in the Leibniz-Mach worldview are that:
\begin{itemize}
\item  there is no such thing as a physical space which is metrical \emph{and} empty; 
\item  physical time is no more than a metaphor for sequential physical process.
\end{itemize}
The quantitative theory was synthesized, using first-principle arguments, to capture these ideas. It has the following structure:
\begin{itemize}
	\item The level surface $\Phi=k$: in the most simple case of the cosmology this represents a three-dimensional spherical surface but, in general, it represents any topological isomorphism of a three-dimensional spherical surface;
	\item The primitive mass function $M(\Phi)$: this represents the amount of material contained within the level surface $\Phi=k$;
	\item The metric structure of physical three-space is projected out of its mass content $M(\Phi)$ according to:
	\begin{eqnarray}
	g_{ab} &=& \frac{1}{8 \pi \Sigma_F }\nabla_a \nabla_b M \equiv \frac{1}{8 \pi \Sigma_F }\left(\frac{\partial^{2}M}{\partial x^{a}\partial x^{b}}-\Gamma_{ab}^{k}\frac{\partial M}{\partial x^{k}}\right),\label{(3)} \nonumber \\
	\Gamma_{ab}^{k} &\equiv&\frac{1}{2}g^{kj}\left(\frac{\partial g_{bj}}{\partial x^a}+\frac{\partial g_{ja}}{\partial x^b}-\frac{\partial g_{ab}}{\partial x^j} \right).\nonumber
	\end{eqnarray} 
	This represents a non-linear differential equation to be solved for $g_{ab}$ in terms of $M$. This is exactly resolvable for the spherical case. No other cases have been attempted. 	Note that if $M(R) \equiv 0$, then there is no metric space;
	\item Physical time is not yet defined, and so there can be no WEP to fix particle trajectories since the WEP is a dynamical principle and relies on a system of clocks;
	\item In place of the WEP we can note that the \emph{shape} of a particle's orbit is also independent of particle properties;
	\item So, using this shape-principle, any particle trajectory minimizes:
	\[
	I(p,q)=\int^q_p{\cal L}\,dt\equiv \int^q_p\sqrt{g_{ij}\dot{x}^{i}\dot{x}^{j}}\,dt \]
	where $t$ is simply an ordering parameter, since physical time is not yet defined. 
	\item The definition of $I(p,q)$ is homogeneous degree zero in the parameter $t$, which means that $I(p,q)$ is invariant under arbitrary monotonic transformations of the $t$-parameter.  This means that the parameter $t$ cannot be physical time without further conditions being attached and, without those conditions (as a standard result), the Euler-Lagrange equations are not a linearly independent set. This is formally identical to the situation in GR, where the same problem is resolved by defining particle proper time. In the present case, once further conditions are imposed to close the EL equations, the closure information has the effect of constraining $t$ to be what is meant by physical time. In practice, the system is closed by specifying that the choice of $t$ must remove the dissipative terms of the EL equations, thereby making the system conservative. We refer to the required condition as the \emph{clock constraint} condition.
\end{itemize}
This model has an equilibrium state which is necessarily associated with a non-trivial fractal $D=2$ distribution of material. 
\subsection{General equations of motion}
There are two distinct cases which arise from the Euler-Lagrange equations:
\subsubsection*{The degenerate case of purely circular motions}
In this case $\dot{R}\equiv0$ and so the radial EL equation
\[
\frac{d}{dt}\left( \frac{\partial {\cal L}}{\partial \dot{R}}\right)-\frac{\partial {\cal L}}{\partial R} = 0 \]
integrates directly to give
\[
{\cal L} = const
\]
leading to
\begin{equation}
{\cal L}^2 = v_0^2 ~~\rightarrow~~ R^2\dot{\theta}^2 = \frac{8\pi\Sigma_Fv_0^2}{A} \equiv \frac{v_0^2}{d_0}\,\left(\frac{4\pi\Sigma_FR^2}{\mathcal{M}} \right) \label{(11)}
\end{equation}
where $d_0$ is a dimensionless parameter, and $v_0$ is a constant with the dimensions of velocity.
This is the basic orbital equation for the degenerate case of conservative MOND. In the Leibniz-Mach interpretation, when the parameter $v_0$ (which has dimensions of \emph{velocity}) is fixed, then (\ref{(11)}) defines the system clock, meaning that one complete orbit defines a unit of time in the Leibniz-Mach worldview. 
\subsubsection*{General case for arbitrary orbits in the spherical system}
The full set of Euler-Lagrange equations for arbitrary orbits (assumed to be equatorial with no loss of generality), together with the \emph{clock constraint} condition which defines what is meant by physical time (discussion of \S\ref{SQM}), can be rearranged as:
\begin{equation}
\mathbf{\ddot{R}} = - \frac{d\mathcal{V}}{dR}\,\mathbf{\hat{R}},
\label{11R} 
\end{equation}
\[
\mathcal{V}=  -\frac{1}{2}\left[4 \,d_0^2 v_0^2\left(\frac{  \mathcal{M}}{4 \pi \Sigma_F } -\frac{ h^2 }{d_0 v_0^2} \right)\left( \frac{  \,\mathcal{M}^2}{R^4 \mathcal{M}' \mathcal{M}'} \right)+\frac{h^2}{R^2} \right], \]
\emph{clock~constraint:}
\[
\frac{1}{2}\left( \dot{R}^2 + R^2 \dot{\theta}^2 \right) \,=\,- \mathcal{V}\]
where the \emph{clock constraint} (which defines the system clock in the Leibniz-Mach worldview) is the condition required to ensure that a disspation term in the original EL system becomes identically zero.
\\\\
It is clear that this choice of the \emph{clock constraint} is also identical to 
the first integral with respect to $t$ of (\ref{11R}) with the condition that the constant of integration which arises from this process is explicitly set to zero. In classical terms, the clock constraint is identical to the energy equation but, of course, in the classical case \lq{time}' is pre-ordained to be Newton's idea of time.
\\\\
At this juncture, it is worth making the obvious point that a classical Newtonian system with an energy equation identical in form to the clock constraint above can only return parabolic orbits so that, at face value, the system (\ref{11R}) cannot possibly reproduce the generality of Newtonian orbits. This potential objection is resolved in \S\ref{NO}, below.
\\\\
To return to the main theme, since angular momentum is trivially conserved by (\ref{11R}), it follows immediately that the total independent content of these equations is, as stated, an equation (the clock constraint) defining elapsed physical time in terms of \emph{process} or \emph{change} and angular momentum conservation. In this way, the proposition, that every particle in the ensemble is a \emph{system clock} is demonstrated.
\\\\
It follows, immediately, that the orbits are given by
\begin{eqnarray}
V_{rot}(R) &=& \frac{h}{R} \nonumber \\
\nonumber \\
V_{rad}^2(R) + V_{rot}^2(R) &=& -2 \mathcal{V}. \label{12R}
\end{eqnarray}
These are the orbits which form the basis of general case conservative MOND in a spherical world.
\subsection{Brief note on general Newtonian orbits} \label{NO}
In order to reproduce the generality of Newtonian orbits, the clock constraint must have the form
\[
\frac{1}{2}\left( \dot{R}^2 + R^2 \dot{\theta}^2\right) = \left( \frac{GM_S}{R} - \omega^2 \right) ,  
\]
where $\omega$ is either real, zero or imaginary for elliptical, parabolic or hyperbolic orbits respectively. In other words, the potential is necessarily defined to be
\[
\mathcal{V} =  - \left( \frac{GM_S}{R} - \omega^2 \right)
\]
exactly, so that distinct orbits have distinct potentials - the potential function becomes a function of dynamical state, so that any Newtonian orbit has its own unique potential state. This means that $\mathcal{M}$ must be defined to ensure that this is the case which means that  it must satisfy (\ref{11R}b) with $\mathcal{V}$ defined as above. 
\section{Some computational details}\label{MassModels}
There are various details which are necessary to reliably reproduce the results of this paper.
\subsection*{The minimization metric}
For minimization problems involving noisy data, it is generally considered best to use a metric based on the $L_1$-norm. So, for a rotation curve with measured velocities $V_{sparc}(R_i),\,i=1..N$, we seek to determine the disk parameters $(R_0, M_0, V_{flat})$ by minimizing:
\[
metric = \sum^N_{i=1} \left| \frac{V_{sparc}(R_i)-V_{theory}(R_i)}{V_{sparc}(R_i)} \right|
\]
with respect to variation in them. This gives far superior results to those arising from use of the $L_2$-norm, for example.
\subsection*{The Nelder-Mead iteration}
Because the data is noisy, it is necessary to run the Nelder-Mead minimization process multiple times for each disk, with randomly generated initial guesses for the RC fitting parameters  $(R_0, M_0, V_{flat})$. In practice, this means running the minimization process (typically) about 2000 times per disk before the results completely settle down.
\section{Least-areas linear regression} \label{LeastAreas}
\begin{figure}[H]
	\centering
	\includegraphics[width=0.6\linewidth]{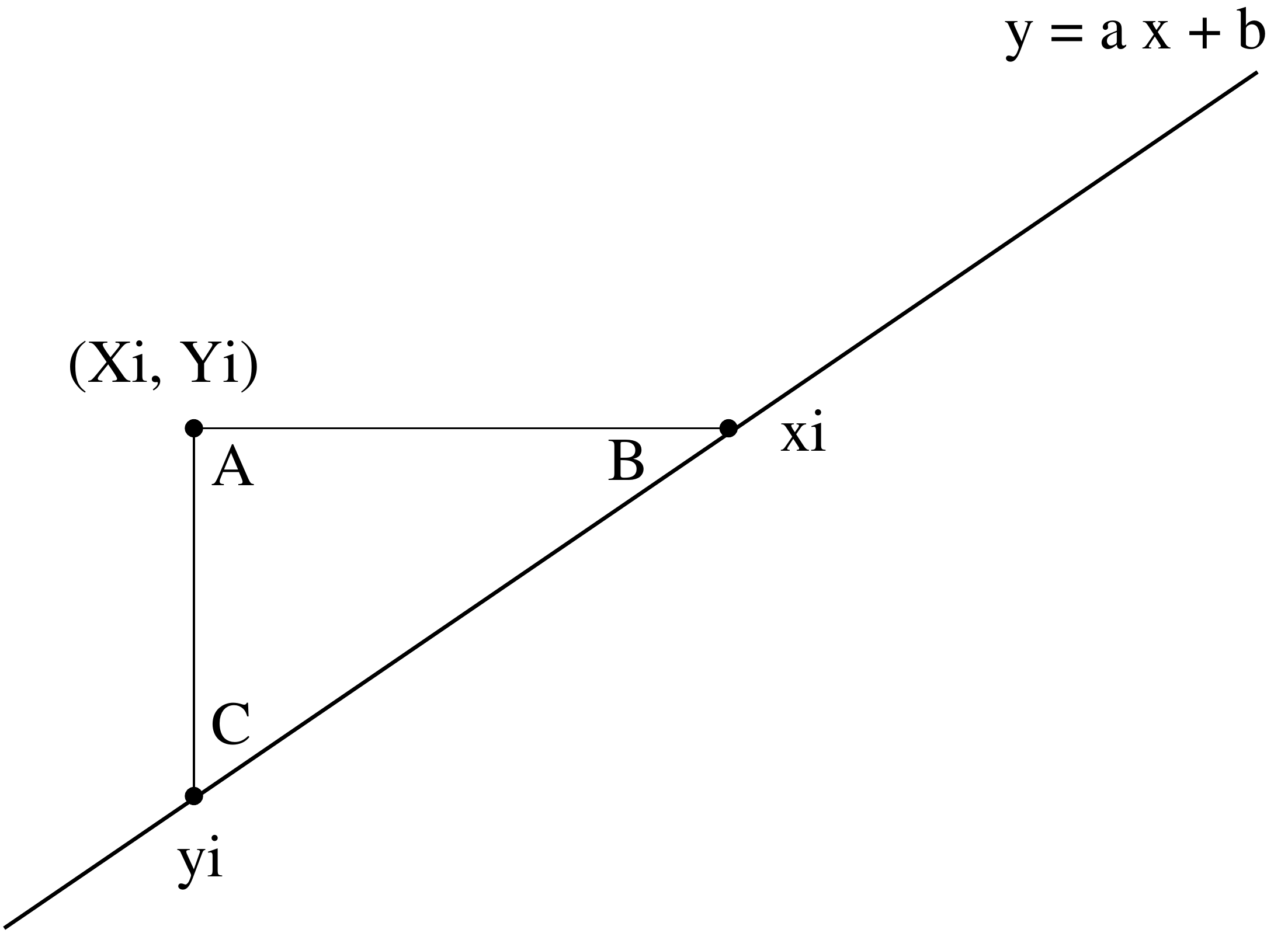}
	\label{Fig:LeastAreas1}
\end{figure}

In the following, the quantity being minimised is independent of how the predictor/response pair is chosen for the regression. The result is a linear model which can be algebraically inverted to give the exact linear model which would also arise from regressing on the interchanged predictor/response pair. So, any inference drawn about the relationship between the predictor and response is independent of how the predictor/response pair is chosen.
\\\\ 
Suppose that we have the data $(X_i, Y_i),~i = 1..N$ to which we fit the model
\begin{equation}
y = A x + B  \label{LA1}
\end{equation}
according to a criterion labelled as \emph{least areas} which we  describe below.
\\\\
From figure \ref{Fig:LeastAreas1}, the area of the triangle shown is given by
\begin{equation}
\Delta_i = -\frac{1}{2}\left(X_i - x_i \right) \left(Y_i - y_i \right), \label{LA2}
\end{equation}
which is always positive, regardless of the position of the point $A$.
From (\ref{LA1}):
\[
x_i =  - \frac{B}{A} + \frac{1}{A }Y_i,~ ~~~y_i = B +  A X_i 
\]
so that (\ref{LA2}) becomes:
\[
\Delta_i = -\frac{1}{2}\left(X_i  + \frac{B}{A} - \frac{1}{A }Y_i  \right) \left(Y_i -B - A X_i \right).
\]
Dropping the numerical factor, the least-area regression model is derived by minimizing
\[
Area = \sum_{i=1}^N \Delta_i = \sum_{i=1}^N \left(X_i  + \frac{B}{A} - \frac{1}{A }Y_i  \right) \left(Y_i -B - A X_i \right)
\] 
with respect to variations in $A$ and $B$. The normal equations are found to be:
\begin{eqnarray}
N B +  A \Sigma X_i  &=& \Sigma Y_i \nonumber \\
\nonumber \\
N B^2 -2 B  \Sigma Y_i - A^2
 \Sigma X_i^2 &=& - \Sigma Y_i^2.
\end{eqnarray}
Hence: 
\[
A = \pm \sqrt{\frac{\left(\Sigma Y_i\right)^2-N \Sigma Y_i^2}{\left(\Sigma X_i\right)^2-N \Sigma X_i^2} }, ~~~
B = \frac{\left( \Sigma Y_i - A \Sigma X_i\right)}{N}
\]
for the result.
\\\\
To see the algebraic invertability property,  fit the model $x = \alpha y + \beta$ to the same data, and we get:
\[
\alpha = \pm \sqrt{\frac{\left( \Sigma X_i\right)^2 - N \Sigma X_i^2}{\left( \Sigma Y_i\right)^2 - N \Sigma Y_i^2}}, ~~~~ \beta = \frac{\Sigma X_i - \alpha \Sigma Y_i}{N}.
\]
Comparing the two models quickly shows that $\alpha = 1/A$ and $\beta = -B/A$ so that $x=\alpha y + \beta$ is the algebraic inverse of $y = A X + B$.
\section{Whole disc scaling law for disk mass} \label{ScalingLaw}
We start with Freeman's Law of (\ref{eqn4f}):
\[
\Sigma_F = \Sigma_0 \left( \frac{V_0}{V_{flat}}\right)^2 = \frac{M_0}{4\pi R_0^2} \left( \frac{V_0}{V_{flat}}\right)^2
\]
from which
\[
\frac{M_0^2}{\mathcal{M}_g(R)}\,\left(\frac{R}{R_0} \right)^2  = M_0 \left( \frac{V_{flat}}{V_0}\right)^2 \frac{4 \pi R^2 \Sigma_F }{\mathcal{M}_g(R)} = M_0 \left( \frac{V_{flat}}{V_0}\right)^2 \frac{\Sigma_F}{\Sigma_R}.
\]
But by (\ref{eqn2}a),
\[
\frac{\Sigma_F}{\Sigma_R} = \left( \frac{V_{rot}(R)}{V_{flat}}\right)^2
\]
so that:
\begin{equation}
\mathcal{M}_g(R) = M_0 \left(\frac{R}{R_0} \right)^2 \left(\frac{V_0}{V_{rot}(R)}\right)^2 ~~~~R \leq R_0. \label{SL1}
\end{equation}
But, from (\ref{eqn5d})
\[
M_{flat}{\rm(theory)} \equiv M_0\,\left(\frac{V_{flat}}{V_0} \right)^2 
\]
from which we can construct an obvious continuation of (\ref{SL1}) as 
\begin{equation}
M(R){\rm (theory)} = M_0\,\left(\frac{V_{rot}(R)}{V_0}\right)^2,~~~~R> R_0,~~~ V_{rot}(R) \leq V_{flat}.
\label{SL2}
\end{equation}
The scaling relations (\ref{SL1}) and (\ref{SL2}) are the whole-disk scaling relations for mass.


\newpage
\begin{thebibliography}{99}
\bibitem[\protect\citeauthoryear{Aguirre}{1999}]{AA1999}	Aguirre, A., 1999, ApJ, 525, 583
\bibitem[\protect\citeauthoryear{Alexander}{1984}]{Alexander1984}Alexander, H.G., \emph{The Leibniz-Clarke Correspondence,}
\bibitem[\protect\citeauthoryear{Baryshev et al}{1995}]{Baryshev1995} Baryshev, Yu V., Sylos Labini, F., Montuori, M., Pietronero, L. 1995 \textit{Vistas in Astronomy} 38, 419
\bibitem[\protect\citeauthoryear{Broadhurst et al}{1990}]{Broadhurst}Broadhurst, T.J., Ellis, R.S., Koo, D.C., Szalay, A.S., 1990, Nat 343, 726
\bibitem[\protect\citeauthoryear{Charlier}{1908}]{Charlier1908}Charlier, C.V.L., 1908, Astronomi och Fysik 4,
1 (Stockholm)
\bibitem[\protect\citeauthoryear{Charlier}{1922}]{Charlier1922}Charlier, C.V.L., 1922, Ark. Mat. Astron. Physik
16, 1 (Stockholm)
\bibitem[\protect\citeauthoryear{Charlier}{1924}]{Charlier1924}Charlier, C.V.L., 1924, Proc. Astron. Soc. Pac. 37, 177

\bibitem[\protect\citeauthoryear{Corasaniti}{2006}]{PSC2006}Corasaniti, P.S., 2006, MNRAS 372, 1, 191-198 



\bibitem[\protect\citeauthoryear{Courteau}{1997}]{SC1997} Courteau S., 1997, AJ, 114, 6, 2402-2427


\bibitem[\protect\citeauthoryear{Da Costa et al}{1994}]{DaCosta} Da Costa, L.N., Geller, M.J., Pellegrini, P.S., Latham, D.W., Fairall, A.P., Marzke, R.O., Willmer, C.N.A., Huchra,
J.P., Calderon, J.H., Ramella, M., Kurtz, M.J., 1994, ApJ 424, L1

 

\bibitem[\protect\citeauthoryear{Dale, Giovanelli \& Haynes}{1997}]{Dale1997} Dale DA, Giovanelli R, Haynes M, 1997
AJ 114 (2): 455-473 

\bibitem[\protect\citeauthoryear{Dale et al}{1998}]{Dale1998} Dale DA, Giovanelli R, Haynes MP, Scodeggio M, Hardy E, Campusano LE, 1998, AJ 115 (2), 418-435


\bibitem[\protect\citeauthoryear{Dale, Giovanelli \& Haynes}{1999}]{Dale1999} Dale DA, Giovanelli R, Haynes MP,  1999, AJ 118 (4), 1468-1488

\bibitem[\protect\citeauthoryear{Dale \& Uson}{2000}]{Dale2000} Dale D.A., Uson JM, 2000, AJ 120 (2), 552-561


\bibitem[\protect\citeauthoryear{Dale et al}{2001}]{Dale2001}
Dale D.A., Giovanelli R, Haynes M.P., Hardy E, Campusano LE, 2001,
AJ 121, 1886-1892 




\bibitem[\protect\citeauthoryear{De Lapparent et al}{1988}]{DeLapparent1988} De Lapparent, V., Geller,M.J., Huchra, J.P., 1988, ApJ 332, 44
\bibitem[\protect\citeauthoryear{De Vaucouleurs}{1970}]{De Vaucouleurs1970} De Vaucouleurs, G., 1970, Sci 167, 1203
\bibitem[\protect\citeauthoryear{Gabrielli \& Sylos Labini}{2001}]{Gabrielli} Gabrielli, A., Sylos Labini, F., 2001, Europhys.
Lett. 54 (3), 286
\bibitem[\protect\citeauthoryear{Giovanelli and Haynes}{1986}]{Giovanelli1986} Giovanelli, R., Haynes, M.P., Chincarini, G.L., 1986, ApJ 300, 77
\bibitem[\protect\citeauthoryear{Haslbauer et al}{2019}]{Kroupa1} Haslbauer, M.; Banik, I.; Kroupa, P.; Grishunin, K., MNRAS, 489, 2, 2634-2651
\bibitem[\protect\citeauthoryear{Huchra et al}{1983}]{Huchra1983} Huchra, J., Davis, M., Latham, D.,Tonry, J., 1983, ApJS 52, 89
\bibitem[\protect\citeauthoryear{Hogg et al}{2005}]{Hogg} Hogg, D.W., Eistenstein, D.J., Blanton M.R., Bahcall N.A, Brinkmann, J., Gunn J.E., Schneider D.P. 2005 ApJ, 624, 54

\bibitem[\protect\citeauthoryear{Joyce, Montuori \& Sylos Labini et al}{1999}]{Joyce} Joyce, M., Montuori, M., Sylos Labini, F., 1999,
Astrophys. J. 514, L5
\bibitem[\protect\citeauthoryear{Labini \& Gabrielli}{2000}]{Labini} Labini, F.S., Gabrielli, A., 2000, \emph{Scaling and fluctuations in galaxy distribution: two tests to probe large scale structures}, astro-ph0008047
\bibitem[\protect\citeauthoryear{Lelli, McGaugh \& Schombert}{2015}]{McGaugh2015} Lelli, F., McGaugh, SS, Schombert, JM., arxiv.org/abs/1512.04543
\bibitem[\protect\citeauthoryear{Mach}{1919}]{Mach1960}Mach, E., 1919, \emph{The Science of Mechanics} - \emph{a Critical and Historical Account of its Development} Open Court, La
Salle, 1960
\bibitem[\protect\citeauthoryear{Martinez et al}{1998}]{Martinez} Martinez, V.J., PonsBorderia, M.J., Moyeed, R.A., Graham, M.J. 1998 \textit{MNRAS} 298, 1212 
\bibitem[\protect\citeauthoryear{Mathewson, Ford \& Buchhorn}{1992}]{MFB1992} Mathewson, D.S., Ford, V.L., Buchhorn, M., 1992, {Astrophys J. Supp.} {81}, 413

\bibitem[\protect\citeauthoryear{Mathewson \& Ford}{1996}]{MF1996} Mathewson, D.S., Ford, V.L., 1996, {Astrophys J. Supp.} {107}, 97

\bibitem[\protect\citeauthoryear{McGaugh}{1995a}]{McGaugh1995a} McGaugh SS, Schombert JM, Bothun GD. 1995a. Astron. J. 109: 2019-2033
\bibitem[\protect\citeauthoryear{McGaugh}{1995b}]{McGaugh1995b} McGaugh SS, Bothun GD, Shombert JM. 1995b. Astron. J. 110: 573-580
\bibitem[\protect\citeauthoryear{McGaugh}{1996}]{McGaugh1996} McGaugh, SS. 1996. MNRAS 280: 337-354
\bibitem[\protect\citeauthoryear{McGaugh}{1998a}]{McGaugh1998a} McGaugh, SS. 1998a. In After the Dark Ages: When Galaxies Were Young, eds. Holt, S.S. \& Smith, E.P.
pp 72-75. AIP (astro-ph/9812328)
\bibitem[\protect\citeauthoryear{McGaugh}{1998b}]{McGaugh1998b} McGaugh, SS, de Blok, WJG. 1998b. Ap. J. 499: 41-65
\bibitem[\protect\citeauthoryear{McGaugh}{1998c}]{McGaugh1998c} McGaugh, SS, de Blok, WJG. 1998c. Ap. J. 499: 66-81
\bibitem[\protect\citeauthoryear{McGaugh}{1999a}]{McGaugh1999a} McGaugh, SS. 1999a. In Galaxy Dynamics, eds. Merritt, D., Sellwood, J.A., Valluri, M., pp. 528-538.
San Francisco: Astron.Soc.Pac (astro-ph/9812327)
\bibitem[\protect\citeauthoryear{McGaugh}{1999b}]{McGaugh1999b} McGaugh, SS. 1999b. Ap. J. Lett. 523: L99-L102
\bibitem[\protect\citeauthoryear{McGaugh}{2000a}]{McGaugh2000a} McGaugh, SS. 2000a. Ap. J. Lett. 541: L33-L36
\bibitem[\protect\citeauthoryear{McGaugh}{2000b}]{McGaugh2000b} McGaugh, SS, Schombert, JM, Bothun, GD, de Blok, WJG. 2000b. Ap. J.
\bibitem[\protect\citeauthoryear{McGaugh}{2001}]{McGaugh2001} McGaugh, SS, Rubin, VC, de Blok, WJG. 2001. Astron. J. 122: 2381-2395
\bibitem[\protect\citeauthoryear{McGaugh}{2016}]{McGaugh2016} McGaugh, SS, Lelli, F. 2016. Phys Rev Lett. 117.
\bibitem[\protect\citeauthoryear{Milgrom}{1983a}]{Milgrom1983a}	Milgrom, M., 1983a, Astrophysical Journal. 270: 365. 
\bibitem[\protect\citeauthoryear{Milgrom}{1983b}]{Milgrom1983b} Milgrom, M., 1983b, Astrophysical Journal. 270: 371
\bibitem[\protect\citeauthoryear{Milgrom}{1983c}]{Milgrom1983c} Milgrom, M. 1983c. Ap. J. 270: 365-370
\bibitem[\protect\citeauthoryear{Milgrom}{1983d}]{Milgrom1983d} Milgrom, M. 1983d. Ap. J. 270: 371-383
\bibitem[\protect\citeauthoryear{Milgrom}{1983e}]{Milgrom1983e} Milgrom, M. 1983e. Ap. J. 270: 384-389
\bibitem[\protect\citeauthoryear{Milgrom}{1984}]{Milgrom1984} Milgrom, M. 1984. Ap. J. 287: 571-576
\bibitem[\protect\citeauthoryear{Milgrom}{1988}]{Milgrom1988} Milgrom, M. 1988. Astron. Astrophys. 202: L9-L12
\bibitem[\protect\citeauthoryear{Milgrom}{1989a}]{Milgrom1989a} Milgrom, M. 1989a. Ap. J. 338: 121-127
\bibitem[\protect\citeauthoryear{Milgrom}{1989b}]{Milgrom1989b} Milgrom, M. 1989b. Astron. Astrophys. 211: 37-40
\bibitem[\protect\citeauthoryear{Milgrom}{1989c}]{Milgrom1989c} Milgrom, M. 1989c, Comments Astrophys. 13: 215-230
\bibitem[\protect\citeauthoryear{Milgrom}{1991}]{Milgrom1991} Milgrom, M. 1991. Ap. J. 367: 490-492
\bibitem[\protect\citeauthoryear{Milgrom}{1994a}]{Milgrom1994a} Milgrom, M. 1994a. AnnalsPhys 229: 384-415
\bibitem[\protect\citeauthoryear{Milgrom}{1994b}]{Milgrom1994b} Milgrom, M. 1994b. Ap. J. 429: 540-544
\bibitem[\protect\citeauthoryear{Milgrom}{1995}]{Milgrom1995} Milgrom, M. 1995. Ap. J. 455: 439-442
\bibitem[\protect\citeauthoryear{Milgrom}{1997a}]{Milgrom1997a} Milgrom, M. 1997a. Phys.Rev.E 56: 1148-1159
\bibitem[\protect\citeauthoryear{Milgrom}{1997b}]{Milgrom1997b} Milgrom, M. 1997b. Ap. J. 478: 7-12
\bibitem[\protect\citeauthoryear{Milgrom}{1998}]{Milgrom1998} Milgrom, M. 1998. Ap. J. Lett. 496. L89-L91
\bibitem[\protect\citeauthoryear{Milgrom}{1999}]{Milgrom1999} Milgrom, M. 1999. Phys. Lett. A 253: 273-279
\bibitem[\protect\citeauthoryear{Milgrom}{2002}]{Milgrom2002} Milgrom, M. 2002. Ap. J. Lett. 577: L75-L77 
\bibitem[\protect\citeauthoryear{Peebles}{1980}]{Peebles1980} Peebles, P.J.E., 1980, The Large Scale Structure of the Universe, Princeton University Press, Princeton, NJ.
\bibitem[\protect\citeauthoryear{Persic \& Salucci}{1995}]{PS1995}
Persic M., Salucci P., 1995, {ApJS}, 99, 501
\bibitem[\protect\citeauthoryear{Pietronero \& Sylos Labini}{2000}]{Pietronero} Pietronero, L., Sylos Labini, F., 2000, Physica
A, (280), 125

%\bibitem[1999a]{RoscoeA}Roscoe D.F., 1999a, Astron. Astrophys.,343, 788-800 
%\bibitem[1999b]{RoscoeB}Roscoe D.F., 1999b, Astron. Astrophys.,343, 697-704 
%\bibitem[$2002$a]{RoscoeC}Roscoe, D.F., $2002$a, Gen. Rel. Grav., 34, 5, 577-602 
%\bibitem[$2002$b]{RoscoeD}Roscoe D.F., $2002$b, Astron. Astrophys.,385, 431-453 
\bibitem[\protect\citeauthoryear{Roscoe}{2002}]{Roscoe2002} Roscoe D.F., 2002, {General Relativity \& Gravitation}, {34}, 577-603
\bibitem[\protect\citeauthoryear{Roscoe}{2020}]{Roscoe2020} Roscoe D.F., A complete Leibniz-Mach cosmology:  arxiv.org/abs/0802.2889
\bibitem[\protect\citeauthoryear{Sanders}{1984}]{Sanders1984} Sanders, RH. 1984. Astron. Astrophys. 136: L21-L23
\bibitem[\protect\citeauthoryear{Sanders}{1986}]{Sanders1986} Sanders, RH. 1986. MNRAS 223: 539-555
\bibitem[\protect\citeauthoryear{Sanders}{1988}]{Sanders1988} Sanders, RH. 1988. MNRAS 235: 105-121
\bibitem[\protect\citeauthoryear{Sanders}{1989}]{Sanders1989} Sanders, RH. 1989. MNRAS 241: 135-151
\bibitem[\protect\citeauthoryear{Sanders}{1990}]{Sanders1990} Sanders, RH. 1990. Astron.Astrophys. Rev. 2: 1-28
\bibitem[\protect\citeauthoryear{Sanders}{1994a}]{Sanders1994a} Sanders, RH. 1994a. Astron. Astrophys. 284: L31-L34
\bibitem[\protect\citeauthoryear{Sanders}{1994b}]{Sanders1994b} Sanders, RH, Begeman, KG, 1994b, MNRAS , 266: 360-366
\bibitem[\protect\citeauthoryear{Sanders}{1996}]{Sanders1996} Sanders, RH. 1996. Ap. J. 473: 117-129
\bibitem[\protect\citeauthoryear{Sanders}{1997}]{Sanders1997} Sanders, RH. 1997. Ap. J. 480: 492-502
\bibitem[\protect\citeauthoryear{Sanders}{1998a}]{Sanders1998a} Sanders, RH. 1998a. MNRAS 296: 1009-1018
\bibitem[\protect\citeauthoryear{Sanders}{1998b}]{Sanders1998b} Sanders, RH, Verheijen MAW. 1998b Ap. J. , 503, 97-108
\bibitem[\protect\citeauthoryear{Sanders}{1999}]{Sanders1999} Sanders, RH. 1999. Ap. J. 512: L23-L26
\bibitem[\protect\citeauthoryear{Sanders}{2000}]{Sanders2000} Sanders, RH. 2000. MNRAS 313: 767-774
\bibitem[\protect\citeauthoryear{Sanders}{2002}]{Sanders2002} Sanders, R., https://arxiv.org/abs/astro-ph/0212293
\bibitem[\protect\citeauthoryear{Sanders \& McGaugh}{2002}]{Sanders2002A} Sanders, R., McGaugh, S., https://arxiv.org/abs/astro-ph/0204521v1
\bibitem[\protect\citeauthoryear{Sanders}{2001}]{Sanders2001} Sanders, RH. 2001. Ap. J. 560: 1-6
\bibitem[\protect\citeauthoryear{Sanders}{2014}]{Sanders2014} Sanders, R. H., 2014, Canadian Journal of Physics. 93 (2): 126
\bibitem[\protect\citeauthoryear{Scaramella et al}{1998}]{Scaramella} Scaramella, R., Guzzo, L., Zamorani, G., Zucca,
E., Balkowski, C., Blanchard, A., Cappi, A., Cayatte, V., Chincarini,
G., Collins, C., Fiorani, A., Maccagni, D., MacGillivray, H., Maurogordato,
S., Merighi, R., Mignoli, M., Proust, D., Ramella, M., Stirpe, G.M.,
Vettolani, G. 1998 \textit{A\&A} 334, 404 
\bibitem[\protect\citeauthoryear{Sylos Labini \& Montuori}{1998}]{SylosLabini} Sylos Labini, F., Montuori, M., 1998, Astron. \& Astrophys., 331, 809
\bibitem[\protect\citeauthoryear{Sylos Labini, Montuori \& Pietronero}{1998}]{SylosLabini1} Sylos Labini, F., Montuori, M., Pietronero, L.,
1998, Phys. Lett., 293, 62
\bibitem[\protect\citeauthoryear{Sylos Labini, Vasilyev \& Baryshev}{2006}]{SylosLabini2} Sylos Labini, F., Vasilyev, N.L., Baryshev,
Y.V., Archiv.Astro.ph/0610938 
\bibitem[\protect\citeauthoryear{Tekhanovich \& Baryshev}{2016}]{Tekhanovich} Tekhanovich D.I.I and Baryshev Yu.V., Archiv.Astro.ph/1610.05206
\bibitem[\protect\citeauthoryear{Vettolani et al}{1993}]{Vettolani} Vettolani, G., et al., 1993, in: Proc. of Schloss Rindberg Workshop: Studying the Universe With Clusters of Galaxies
\bibitem[\protect\citeauthoryear{Wittenburg et al}{2020}]{Kroupa} Wittenburg, N.; Kroupa, P.; Famaey, B., Ap. J., 890, 2
\bibitem[\protect\citeauthoryear{Wu, Lahav \& Rees}{1999}]{Wu} Wu, K.K.S., Lahav, O., Rees, M.J., 1999, Nature
397, 225
\end{thebibliography}
\end{document}